\documentclass[smallextended]{svjour3}     

\usepackage{graphicx}

\usepackage{natbib}
\usepackage{epsfig}
\usepackage{psfrag}

\usepackage{lscape}
\usepackage{float}
\usepackage{rotating}
\usepackage{subfigure}

\usepackage{amssymb}
\usepackage{amsmath}

\usepackage{multirow}

\def\deg{\ensuremath{^{\circ}}}
\def\<{\ensuremath{<}}
\def\>{\ensuremath{>}}

\begin{document}
\journalname{Experimental Astronomy}
\title{The Australia Telescope 20\,GHz Survey: Hardware, Observing Strategy, and Scanning Survey Catalog
}
\subtitle{}

\titlerunning{The AT20G: Scanning Survey Catalog}        

\author{
Paul J. Hancock$^{1,\star} $ \and
Paul Roberts$^2$ \and
Michael J. Kesteven$^2$ \and
Ronald D. Ekers$^2$ \and
Elaine M. Sadler$^1$ \and
Tara Murphy$^{1,3}$ \and
Marcella Massardi$^4$ \and
Roberto Ricci$^5$ \and
Mark Calabretta$^2$ \and
Gianfranco de Zotti$^{4,6}$ \and
Philip G. Edwards$^2$ \and
Jennifer A. Ekers$^2$ \and
Carole A. Jackson$^2$ \and
Mark Leach$^2$ \and
Chris Phillips$^2$ \and
Robert J. Sault$^7$ \and
Lister Staveley-Smith$^8$ \and
Ravi Subrahmanyan$^{2,9}$ \and
Mark A. Walker$^{10}$ \and
Warwick E. Wilson$^2$
}

\authorrunning{Hancock et al.} 

\institute{$\star$ - \email{Paul.Hancock@sydney.edu.au}\\
1 - Sydney Institute for Astronomy (SIfA), School of Physics, The University of Sydney, NSW 2006, Australia\\
2 - CSIRO Astronomy and Space Science (CASS), PO Box 76, Epping, NSW 1710, Australia\\
3 - School of Information Technologies, The University of Sydney, NSW 2006, Australia\\
4 - INAF-Osservatorio Astronomico di Padova, Vicolo dell'Osservatorio 5, I-35122 Padova, Italy \\
5 - INAF-Istituto di Radioastronomia, Bologna, Via P. Gobetti, 101, 40129 Bologna, Italy\\
6 - SISSA, Via Bonomea 265, I-34136 Trieste, Italy\\
7 - School of Physics, The University of Melbourne, Victoria 3010, Australia\\
8 - School of Physics, The Univeristy of Western Australia, 35 Stirling Highway Crawley, WA 6009, Australia\\
9 - Raman Research Institute, C. V. Raman Avenue, Sadashivanagar, Bangalore 560080, India\\
10 - Manly Astrophysics, 3/22 Cliff St, Manly 2095, Australia\\
}

\date{Received: date / Accepted: date}

\maketitle

\begin{abstract}
The Australia Telescope 20GHz (AT20G) survey is a large area (2$\pi$ sr), sensitive (40\,mJy), high frequency (20\,GHz) survey of the southern sky. The survey was conducted in two parts: an initial fast scanning survey, and a series of more accurate follow-up observations. The follow-up survey catalog has been presented by \citet{murphy_australia_2010}. In this paper we discuss the hardware setup and scanning survey strategy as well as the production of the scanning survey catalog.

\end{abstract}

\section{Introduction}
The radio-source population above 5\,GHz has not been well studied, in most part due to the lack of large-area surveys. Large aperture radio telescopes typically have fields of view of a few arc-minutes at high frequencies, making it extremely time-consuming to carry out such surveys. 
The high-frequency properties of radio sources are important to our understanding of the nature and evolution of both radio galaxies and the universe itself. Radio galaxy evolution can be explored via AGN, whilst the nature and evolution of the universe can be explored via studies of the cosmic microwave background (CMB). High resolution radio surveys are crucial for the removal of foreground radio sources \citep{de_Zotti_planck_1999} for CMB missions such as Planck \citep{tauber_planck_2005}, Fermi \citep{ritz_fermi_2009} and the South Pole Telescope \citep{stark_south_1998}, where foreground point sources contaminate the measured CMB anisotropy at angular scales of less than 30 arc-minutes.

A  Wide-Band Analog Correlator (WBAC) was built as a prototype for the Taiwanese Array for Microwave Background Anisotropy \citep[AMiBA][]{lo_amiba:_2001}. This was not used in the production version of the AMiBA project, due to the complexity needed to remove 1/f noise in a large correlator, but it was realised that the WBAC could be used on the ATCA for a large area high frequency survey on a timescale that was not prohibitive. Since the time needed to survey a region of sky to a particular flux density depends both on the size of the primary beam and the integration time per pointing, the 8\,GHz bandwidth offered by the WBAC corresponds to a 32 fold increase in observing speed for a survey of the southern sky, as compared to the 256\,MHz bandwidth available on the ATCA in 2004.

\section{AT20G Hardware}
\label{sec:hardware}
The AT20G survey was conducted using three of the six antennas that make up the ATCA, along with a custom Wide-Band Analog Correlator (WBAC). In this section we describe the antenna and correlator design.

\subsection{ATCA telescopes}
To minimize the selection against extended sources in the survey, the most compact ATCA configuration was used. Antennas 2, 3, and 4 of the ATCA were situated on stations W102, W104, and W106, giving two 30.6m (2-3 and 3-4) baselines and a single 61.2m (2-4) baseline. The antenna design and construction is outlined in \citet{frater_australia_1992}, and the 20\,GHz upgrade to the Compact Array is discussed in \citet{wong_millimetre_2002} and \citet{moorey_cryogenically_2008}.

\subsection{WBAC design}
\begin{figure}[hbt]
 \epsfig{file=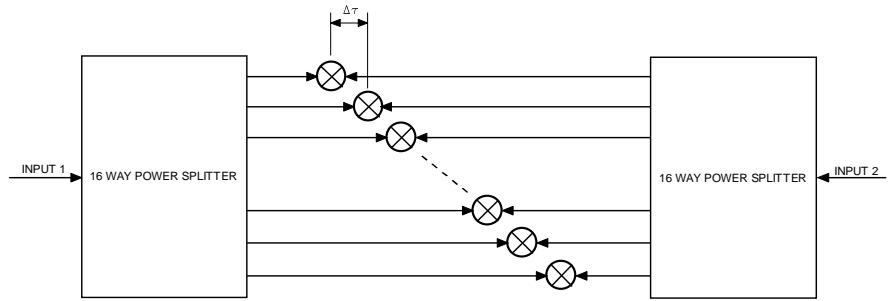, width=\linewidth}
 \caption[WBAC architecture for a single baseline]{Correlator architecture for a single baseline. The multipliers are located at the positions marked by $\otimes$ and the path length differences correspond to the delay spacing of $\Delta\tau$. The complete WBAC consists of three such correlators.}
 \label{fig:corr_arch}
\end{figure}

In 2002/2003 the WBAC was used in a pilot survey as a two element 3.4\,GHz bandwidth interferometer at 18\,GHz. The pilot survey covered approximately 1200\,deg$^2$ of the sky between $-59\deg>\delta> -71\deg$, for further details see \citet{ricci_first_2004}. The wide-band analogue correlator was adapted and expanded to produce a 3 element 8\,GHz bandwidth, dual orthogonal polarization interferometer with a central frequency of 20\,GHz, which was used for the main AT20G survey.

\subsubsection{Assembly}
\label{sec:wbacassembly}
The WBAC is a lag correlator in which the baseline correlation functions are calculated firstly, and the cross power spectrum secondly. The operation of a lag correlator, and comparison against an FX architecture are discussed in \citet{holler_6-12_2007}. He we restrict our discussion to details specific to the WBAC. The lag correlator is constructed from an array of 16 multipliers on each of the three baselines, with precise transmission line delays providing the required delay difference for each lag. The architecture chosen is shown in Figure \ref{fig:corr_arch} and comprises two wide bandwidth power splitters to replicate the input signals, which are then arranged as two parallel counter-propagating sets of signals. The correlators were designed to process a 0-12\,GHz input band to avoid having to perform a final down-conversion of the 4-12\,GHz IF band to baseband. At the intersection of each corresponding pair of transmission lines is placed a multiplier. To facilitate assembly and simplify any required rework the multipliers are mounted on small second level microwave circuit boards which are inserted into cutouts milled in the correlator boards at the required positions.

\begin{figure}[hbt]
 \centering
 \epsfig{file=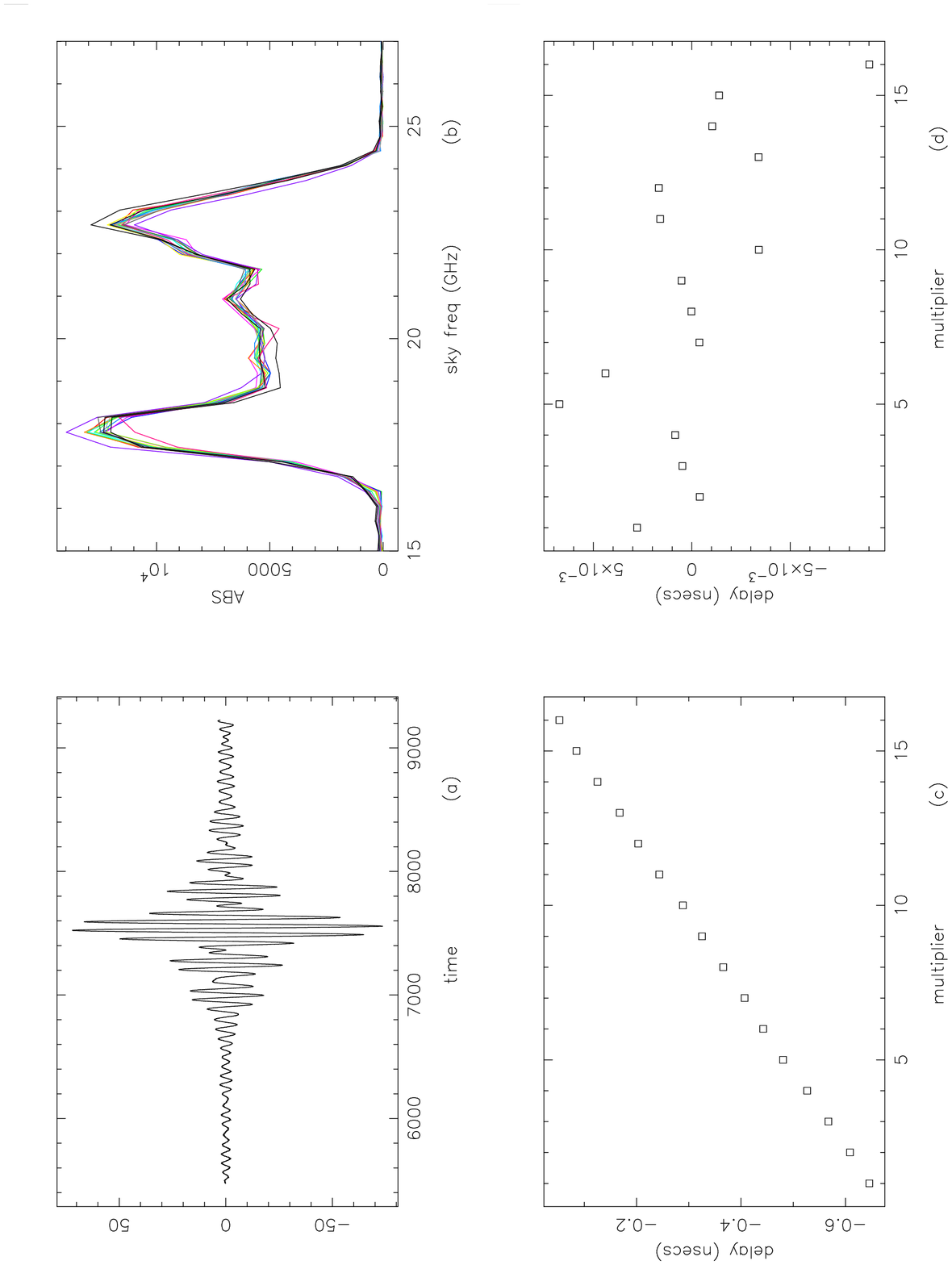,bb=310 0 545 705 ,height=\linewidth, angle=-90,clip=}
 \caption[The transform phases of the WBAC for the primary calibrator 1921$-$293]{{\bf Left:}The transform phases for the WBAC with reference to an arbitrary time for the primary calibrator 1921$-$293. The spacing between the correlator lag channels is of the order $4.2\times10^{-2}$\,nsec. {\bf Right:} The deviation from a linear fit to the transform phases. The largest error is $0.7\times10^{-2}$\,nsec with a typical value of $0.2\times10^{-2}$\,nsec.}
 \label{fig:delaytransform}
\end{figure}

The separation of the lag channels, $\Delta \tau$, is determined by the displacement of the multiplier boards as shown in figure \ref{fig:corr_arch}, and is designed to be a half wavelength at 12\,GHz (ie Nyquist), which is a delay of $4.2\times10^{-2}$\,nsec. The sampling in time delay is not perfectly uniform due to manufacturing and material effects, and may also have a small frequency dependence. This deviation from ideal Nyquist sampling is stable in time and can be accurately measured so the spectrometer's performance is not significantly affected. Figure \ref{fig:delaytransform} shows the transform phases of the WBAC as a function of channel number, for the primary calibrator 1921$-$293. The deviation from a linear fit to the phases is also shown, and is typically $0.2\times10^{-2}$\,nsec, which is less than 5\%. The non ideal sampling is overcome by additional data processing steps as outlined in section \ref{sec:imaging}. In a digital correlator the delay spacings are able to be adjusted so that the delay spacings are all equal and would result in a flat line at 0\,nsec for the right hand plot of \ref{fig:delaytransform}.

\subsubsection{Data Acquisition}
Timing and data acquisition of the correlator system is controlled from a Linux based computer, running a modified version of the standard Compact Array correlator control software. High accuracy timing signals are taken from the observatory's 5\,MHz reference signal which is derived from a hydrogen maser. The data from the ATCA antennas are transported via an optical fiber to the control building where it is then converted to an electric signal by an optical receiver. The three antennas each have a phase switch in the signal path just after the optical receivers. The phase switching is designed to be mutually orthogonal and take the form of a 350 kHz square wave and two 700 kHz square waves.
The signals are correlated and sent to an analogue to digital converter (ADC) where they are accumulated for an integration cycle before being sent to the control computer where the data are archived.

\subsubsection{WBAC Performance}
Stability of the WBAC system was measured by observing the variance in the integrated spectrum as a function of time using a laboratory noise source and amplifier providing nominal input levels. This variance reduced in accordance with the radiometer equation, as the inverse square root of the integration time. This measurement was only taken over 10 minutes but was adequate to verify stability for the current transit instrument, which operates with integration time less than 100\,ms. The stability of the WBAC is important for the flux calibration of the survey observations as it allows for the use of a single bandpass and flux calibrator.

\begin{figure}[hbt]
 \centering
 \epsfig{file=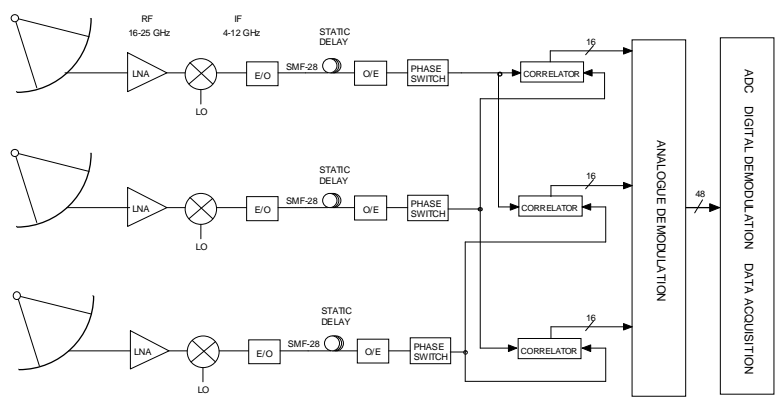, width=\linewidth}
 \caption{Schematic representation of the combined antenna and analog correlator system.}
 \label{fig:sys_diag}
\end{figure}

\subsection{Combined System}
Figure \ref{fig:sys_diag} shows a block diagram of the analogue correlation system. The system is comprised of three main parts: The IF transmission system that transports the down-converted received signals to the central site; the correlators; and the data acquisition and processing system. The ATCA analogue correlator lacks a variable delay compensator so it can be used only with transit interferometers.The static delays are adjusted so that the array is directed towards the meridian with the antennas in a East-West linear array.

\subsubsection{Performance}
Once the WBAC was connected to the telescopes further characterisation and alignment of the system was performed. We adjusted the static delays to maximize the fringe amplitudes at the time a point source passed through the primary beam.

\begin{figure}[hbt]
 \epsfig{file=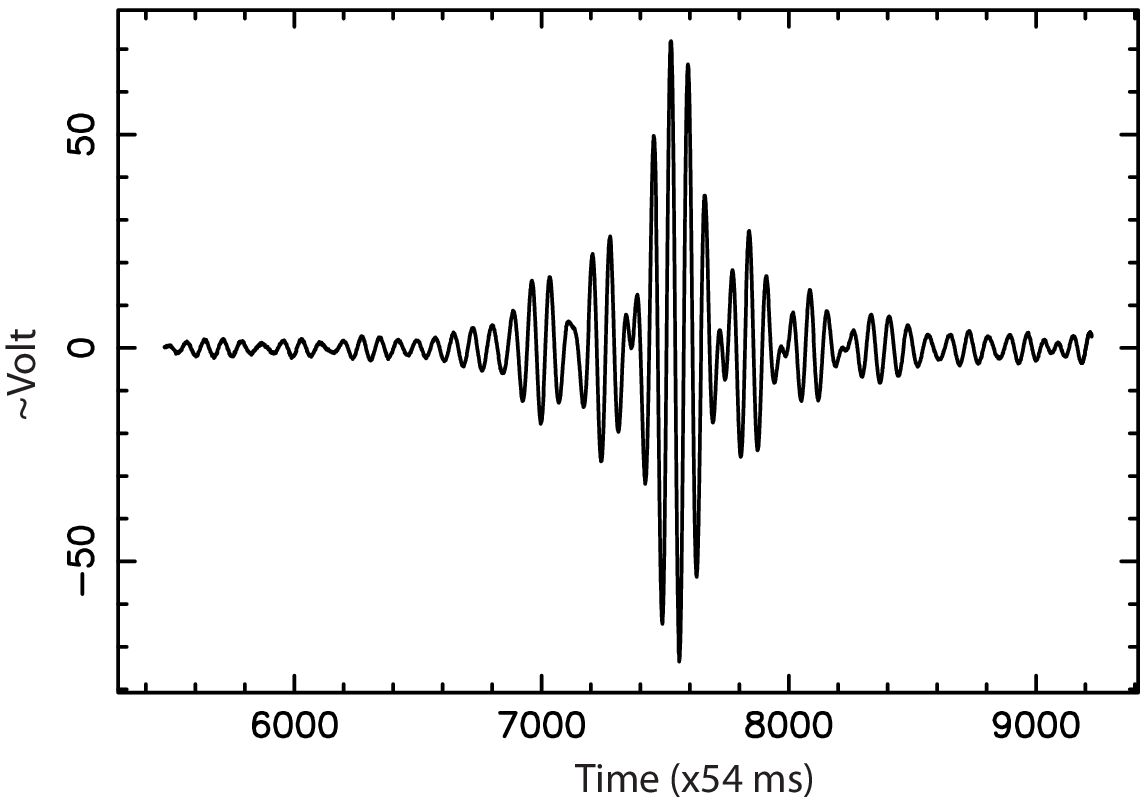,bb= 0 0 330 235, width=0.48\linewidth,clip=}
 \epsfig{file=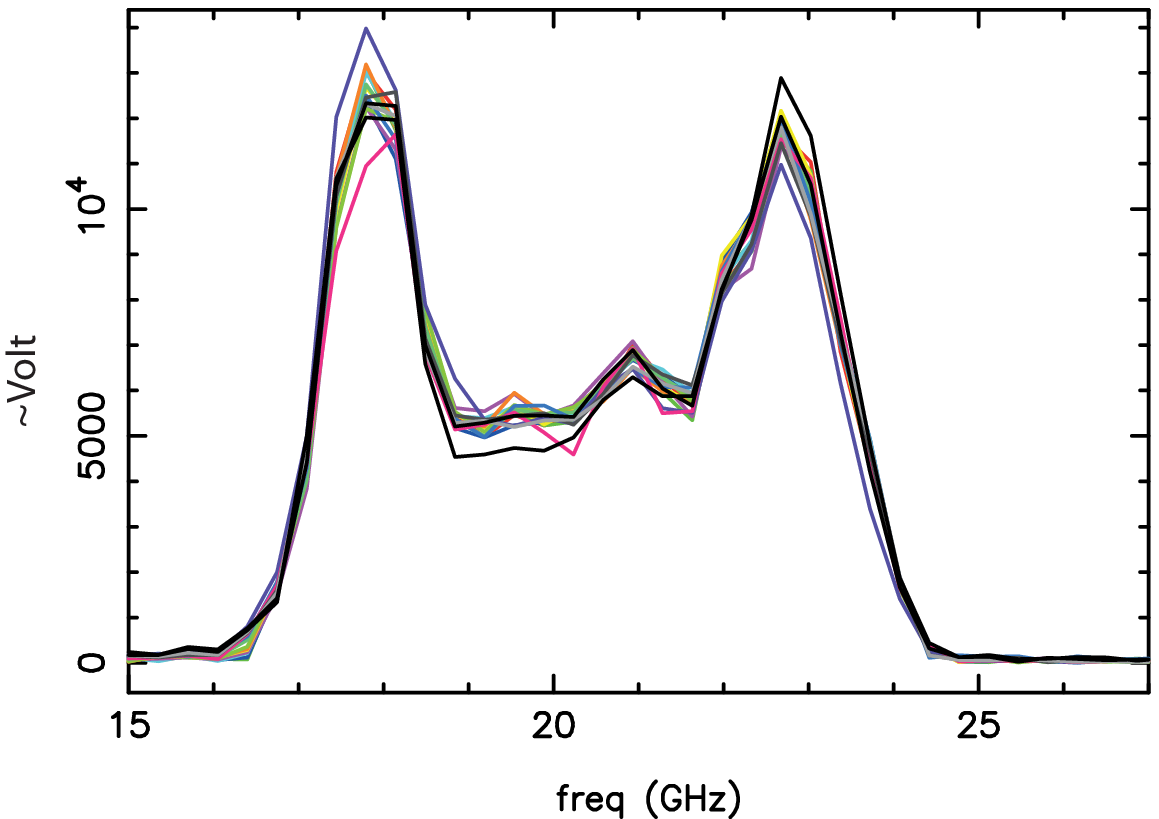,bb=0 0 330 235, width=0.48\linewidth,clip=}
 \caption[ATCA+WBAC system response for AT calibrator 1921$-$293]{ATCA+WBAC system response for AT calibrator 1921$-$293. {\bf Left:} Delay beam fringe pattern for a single lag channel on baseline $2-3$. {\bf Right:} A Fourier transform of a lag channel gives the system bandpass. All 16 channels are superimposed.}
 \label{fig:sys_res}
\end{figure}

The fringe pattern produced by each lag channel can be Fourier inverted to give the entire system frequency response for each correlator channel. From this data each of the lags can be calibrated. Figure \ref{fig:sys_res}(Left) shows the recorded fringe pattern for a single lag within the correlator as a calibrator source was observed. In figure \ref{fig:sys_res}(Right) the Fourier inverted system band-passes for all 16 lags in the correlator are superimposed on a linear scale. The characteristic saddle shape is dominated by the gain curve of the 4-12 GHz amplifiers used in the IF transmission system.

The fundamental determinant of performance is the sensitivity achieved with the system. Observations with the 8 GHz, three baseline system were begun in September 2004. During these observations an average rms sensitivity of approximately 10 mJy, using 80 ms integration time, was achieved.

\section{Scanning strategy}
\begin{figure}
 \epsfig{file=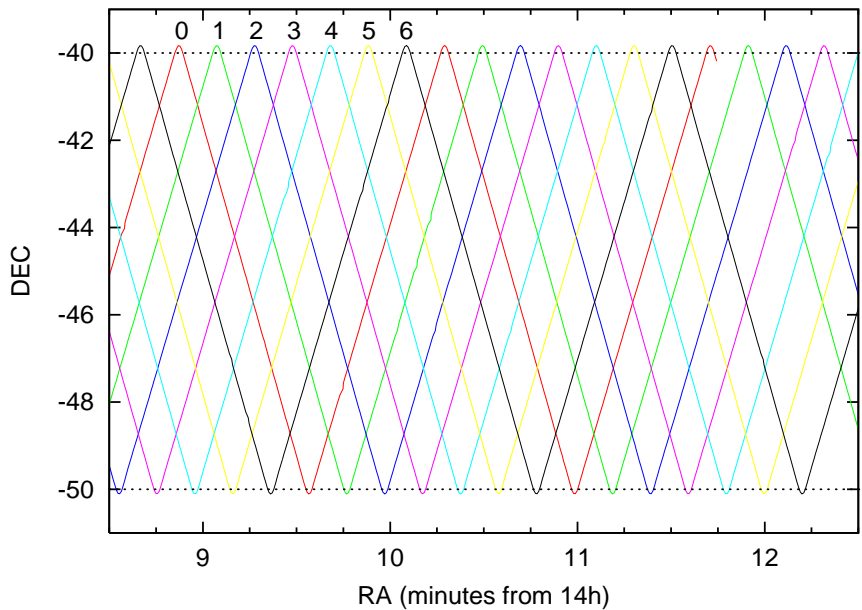, height=\linewidth, angle=-90}
 \caption{An example of a scanning interleave pattern for $-40\deg\geq\delta\geq -50\deg$. A total of 7 interleaves were required to achieve an interleave spacing equal to the primary beam FWHM of 2.3 \,arcmin. Data that were flagged from interleave 0 (red) near $14^h12^m$, is not a terminal problem as the same area of sky is covered once by observations scanning in the opposite direction.}
 \label{fig:interleave}
\end{figure}

As it was not possible to insert the geometric delays necessary to track a point on the sky at 8\,GHz bandwidth, the survey was intentionally restricted to meridian observations. By scanning the telescopes close to the maximum rate of $\sim$15\deg min$^{-1}$ along the meridian, the natural rotation of the earth allowed the observation of interleaving tracks across the sky. The position of the antennas was recorded every $\sim$50ms. This gave a 0.8\,arcmin spacing between observations along each track. The telescope slew speed and integration times were fine tuned so that the tracks would produce closed paths (over a 24 hour period), with each subsequent path being one half-power beam width ($2.3$\,arcmin) from the previous scan. Figure \ref{fig:interleave} shows the interleaved tracks that are formed when data from multiple days are combined. The closed path allows for the distinction between tracks (interleaves), which is important to ensure that any data which are missed or flagged due to bad weather or instrumental failure is easily repeated. The separation of the tracks by a beam width means that each point on the sky falls within the FWHM of the primary beam of the telescope. Sky locations on the interior of the scan region are covered twice each by this scanning strategy, once in a northward track and once in a (different) southward track. Towards the boundaries of the declination band the coverage becomes less uniform and points are either covered only once, at the extreme edges, or twice with a single track.

The survey mode observations for any given declination band of the sky were carried out contiguously, with all the observations for a particular region of the sky being completed within a single observing period of typically 10 days. Observing in this way takes advantage of the stability of the correlators. Table \ref{tab:obsdates} shows when the observations for each declination band were taken. To make the best use of the available time, any observations that needed to be redone were collected and generally carried out towards the end of the allocated observing time. For small amount of bad data that occurred towards the beginning or end of a particular interleave, the patching was done during the change over to the next day's observation. Due to our meridian scanning technique and the randomness of bad weather or equipment failure, it was not always possible to recover all the data which were lost or flagged. When patching, preference was given to regions with missing data, and then to those with large amounts of noise or bad weather.

\begin{table}[hbt]
 \centering
 \begin{tabular}{rr}
\hline
 Declination Band & Epoch \\
\hline
$-30\deg\geq\delta\geq -40\deg$  & 11 Aug - 31 Aug 2004 \\
$-40\deg\geq\delta\geq -50\deg$  & 20 Aug - 31 Aug 2004 \\
$-50\deg\geq\delta\geq -60\deg$  &  9 Sep -  2 Oct 2005 \\
$-60\deg\geq\delta\geq -70\deg$  & 23 Sep -  2 Oct 2005 \\
$-70\deg\geq\delta\geq -80\deg$  & 16 Sep - 20 Sep 2005 \\
$^\dagger$ $-80\deg\geq\delta\geq -90\deg$  & 20 \& 29 Sep 2005 \\
$-15\deg\geq\delta\geq -30\deg$  & 16 Aug -  3 Sep 2006 \\
$  0\deg\geq\delta\geq -15\deg$  & 23 Aug -  9 Sep 2007 \\
$^\dagger$ $-85\deg\geq\delta\geq -90\deg$  & 7 \& 9 Sep 2007 \\

\hline
 \end{tabular}
 \caption{The observation dates for the scanning observations for each of the declination bands. Overlapping dates are due to patching data being observed at the end of each year's observations. $^\dagger$These regions include the south celestial pole and as such, required a separate program to produce the maps.}
 \label{tab:obsdates}
\end{table}

\section{Survey Calibration}
\label{sec:calibration}
As mentioned in section \ref{sec:wbacassembly}, the WBAC has lag channel spacings that are not regular. In the absence of noise, the irregular lag channel spacing will not affect the recovery of the observed spectrum. With noise present, noise correlation between lag channels will increase as the spacing diverges from Nyquist. To avoid correlated noise signals, the calibration and data reduction of the scanning interleaves was carried out in the lag domain. Calibration of the lag response can be done using monochromatic signals \citep{harris_wideband_2001}, or {\em in situ} using astronomical sources \citep{holler_6-12_2007}. The AT20G survey had already adopted the approach independently described by \citet{holler_6-12_2007}.

It is necessary to be able to deduce the location of a source in delay space from the signal distribution of the 16 lag channels of the correlator. If the calibration is correct over the three correlators then we will derive a common delay when a source is within the antenna beam. The source's position on the sky is derived from the delay, the antenna elevation, and the UT of the observation. Once the correlators are calibrated this process can be inverted: that is, given a record of 16 lags over 3 baselines, it is possible to estimate, for any delay, the amplitude of a source at a particular location which is consistent with the record. The key to this process is the tracking calibration which gives the signatures for all the delays within the correlator's range. A source on the equator observed with a 30m baseline will have 75 recorded samples in the time it takes for the peak correlation to advance from one multiplier to the next.

\subsection{Overview}
\label{sec:calibrationoverview}
\begin{figure}[hbt]
 \centering
 \epsfig{file=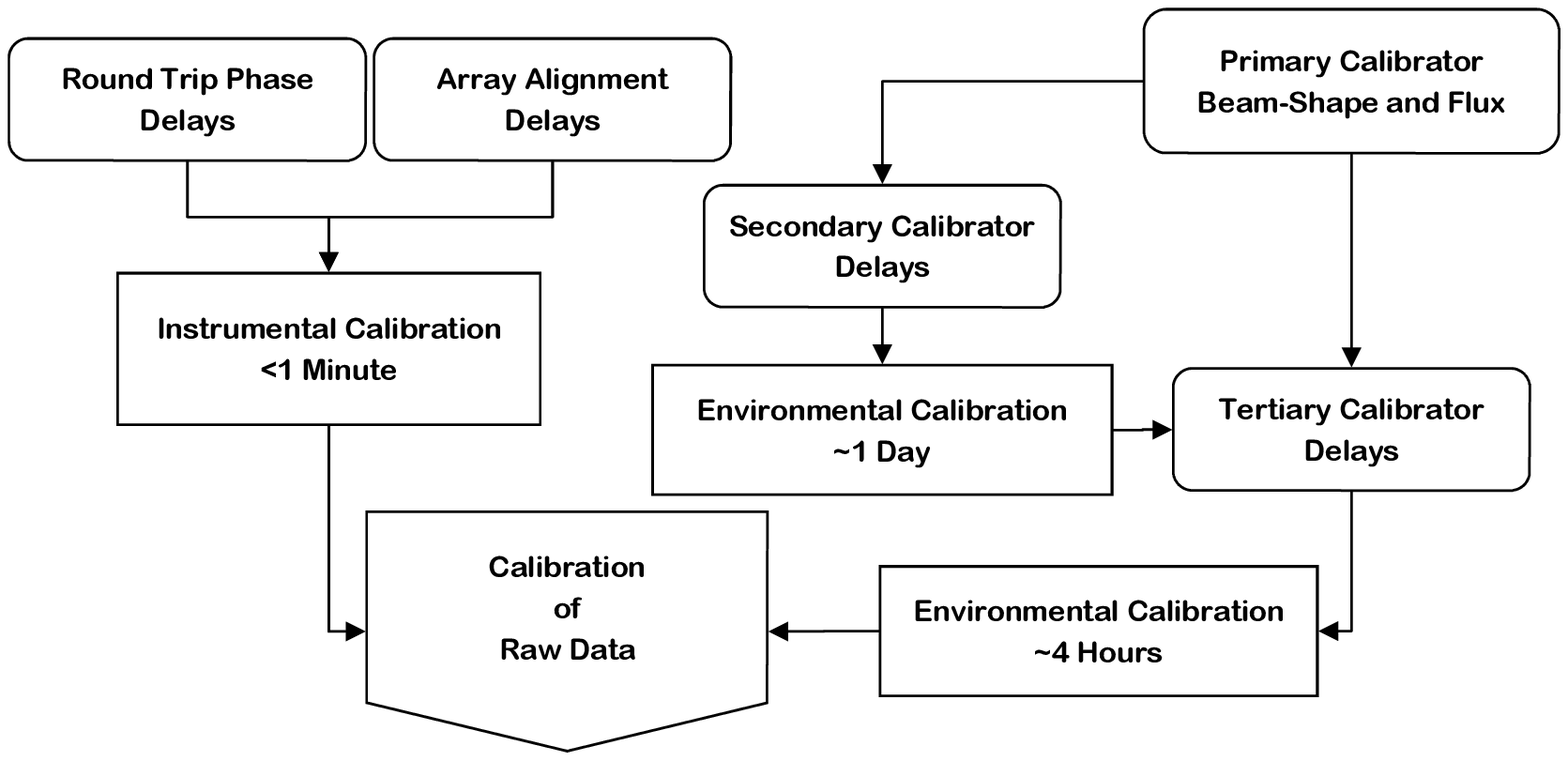,bb=55 540 535 775,width=0.8\linewidth,clip=}
 \caption{An overview of the various delay and flux calibration contributions.}
 \label{fig:calibration}
\end{figure}

The calibration scheme used for the scanning interleave data has five main components, two of which are instrumental and three of which are environmental. The instrumental calibration components are the round trip phase information, and the array alignment correction. These are measured and corrected for on timescales of less than a minute. The environmental calibration components use astronomical sources to detect variations in the correlator delay response and make the required adjustments. Figure \ref{fig:calibration} shows an overview of the calibration contributions.

As there were no standard programs available that could cope with the specific setup of the WBAC, it was necessary to create a suite of purpose built programs. The programs were written in FORTRAN 77\footnote{and later upgraded to FORTRAN 90} and implemented as tasks within the {\sc miriad} environment \citep{sault_retrospective_1995}. The software design and implementation is detailed in an upcoming ATNF memo.

\subsection{Timing Calibration}
\begin{figure}[hbt]
 \centering
 \epsfig{file=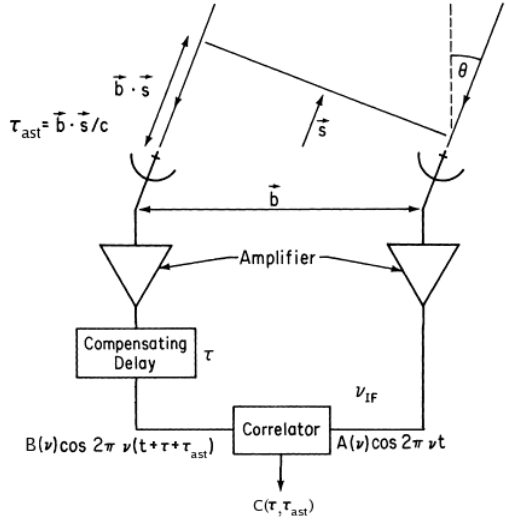, height=0.6\linewidth}
 \caption[Schematic of a single baseline, single channel interferometer]{Schematic of a single baseline, single channel east--west interferometer. The astronomical delay $\tau_{ast}$ is a function of meridian angle $\theta$, and baseline b. Each correlator within the WBAC computes $C(\tau_{ast})$ for 16 different delays ($\tau$) resulting in the function $C(\tau,\tau_{ast})$. Figure adapted from \citet{thompson_fundamentals_1999}}
 \label{fig:interferometer}
\end{figure}

The WBAC consists of 3 correlators, each correlating 16 lag channels for a single baseline. The output from each of the correlators' lag channels can be represented by a single baseline single channel interferometer as depicted in figure \ref{fig:interferometer}. The following is a simplified derivation of the received signal from each of the correlators, where noise terms, phase differences, and local oscillator mixing effects have been neglected for clarity.

The signal received from each of the amplifiers can be written as:
\[ V_{A} = A(\nu) \cos[2\pi \nu t]\]
\[ V_{B} = B(\nu) \cos[2\pi \nu(t+\tau+\tau_{ast})]\]
where: $A(\nu)$ and $B(\nu)$ describe the voltage bandpass of the amplifiers, the data transmission path from antenna to the correlator, and the gain equalisation units at the correlator, $\tau$ is the compensating delay inserted by the correlator, and $\tau_{ast}$ is the astronomical delay:
\[ \tau_{ast} = \frac {b}{c} \sin (\theta) \]
for an East-West baseline of length b observing a source at meridian angle $\theta$.

The correlator receives the two signals and multiplies them to form:

\[ V_{A}V_{B} = M(\nu)A(\nu)B(\nu) \cos[2\pi \nu t] \cos[2\pi \nu(t+\tau+\tau_{ast})]\]
\[ =  M(\nu)A(\nu)B(\nu)\frac{1}{2}\{\cos[2\pi\nu(2t+\tau+\tau_{ast})] +\cos[2\pi\nu(\tau+\tau_{ast})] \}\]

where $M(\nu)$ is the bandpass of the multiplier and the result been separated into a high and low frequency component. A low pass filter is used to remove the high frequency component with frequency $4\pi\nu t$, leaving only the low frequency component which is then integrated to give the correlation function:

\[ C(\tau, \tau_{ast}) = \int^{\nu_c +\frac{\Delta\nu}{2}}_{\nu_c -\frac{\Delta\nu}{2}} M(\nu)A(\nu)B(\nu) \cos[2\pi \nu(\tau+\tau_{ast})] d\nu \]

where $\nu_c$ is the low pass filter's central frequency and $\Delta\nu$ is its bandwidth.

\begin{figure}
\centering
\epsfig{file=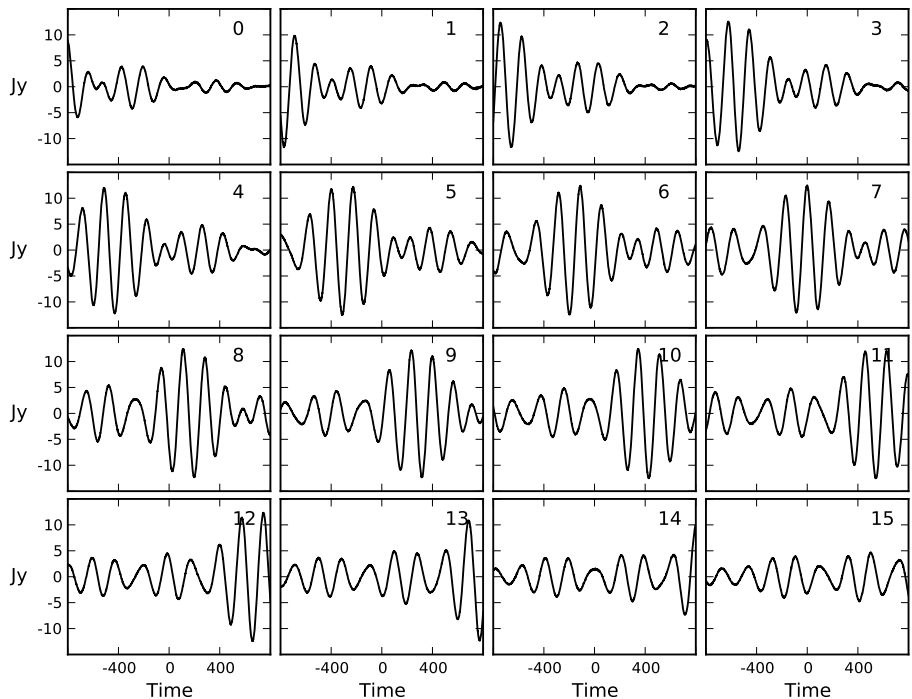, bb=70 200 540 575, width=0.98\linewidth}
\caption{The correlator response to the calibrator 1921$-$293 on baseline 2$-$3. All 16 lag channels are shown and the channel number is indicated in the upper right of each plot. The abscissa is time since transit in units of 54ms. The ordinate is un-calibrated flux in units of Jy.}
\label{fig:lag_response}
\end{figure}

The basic observable is the set \{$C(\tau,\tau_{ast})$\}. Primary calibrator observations are done in a tracking mode where by the calibrator is tracked by the telescopes as it transits the meridian. During this time the set $C(\tau,\tau_{ast})$ is recorded for range of delays $\tau_{ast}$ that is large enough to encompass the synthesised beam response as is demonstrated in Figure \ref{fig:lag_response}. Having the antennas track the calibrator allows the primary beam response of the antennas to be disentangled from the synthesised beam response of the correlator. Once $C(\tau,\tau_{ast})$ has been measured, $A(\nu)B(\nu)M(\nu)$ can be derived via Fourier inversion, and the correlator response can be treated as given in the above equation. The separation between lag channels ($\Delta\tau$) is $\lambda/2$ where $\lambda=24$mm and corresponds to 85\,arcsec in RA for a 30m baseline. The separation in $\tau_{ast}$ is the integration period of 54\,ms and represents $~1$\,arcsec in RA.

In addition to the primary calibrator observation, timing adjustments are required for environmental, and instrumental effects that vary on time scales of minutes to hours. The main environmental factor is the phase difference between the signals from two telescopes due to their slightly different transmission paths. The ATCA is equipped with a round trip phase tracking system that measures the delays introduced into the telescope to correlator transmission path due to thermal effects within the conduits containing the optic fibers. The round trip phase is sensitive to diurnal changes in temperature within the conduits as well as the airconditioning cycles within the telescopes and correlator room. The main instrumental factor is the array alignment. The ATCA array deviates slightly from true East-West, requiring an elevation dependent correction to the timing solution. 

Secondary and tertiary calibrators are used to refine the timing corrections that are not otherwise captured by the ATCA monitoring system. The number of calibrators within the ATCA calibrator catalog is high enough that, on average, 3-4 daily calibrators will fall within the primary beam of the array during the normal survey scanning observations. Since each part of the sky is covered by two separate interleaves each of these secondary calibrators appears in at least two different interleaves and thus provides a way to calibrate the inter-scan timing solutions. To increase the number of calibrators available, strong ($S_{20}\> 100$\,mJy, SNR \> 5) sources found within the calibrated scanning observations were used to self calibrate the timing, by requiring phase closure between the three baselines. The number of available calibrators was increased from around 500 to nearly 3000 over the entire sky using this bootstrapping process.

Once the follow-up observations of the AT20G survey had been completed and a list of confirmed sources had been created, a final round of calibration was done using the improved positions and fluxes of all the sources with SNR$\ge5$. It is from this final round of calibration that the catalog described in section \ref{sec:catalog} was created.

\subsection{Flux Calibration}
\label{sec:fluxcalibration}
In the calibration scheme outlined in the previous sections, the primary focus is on timing calibration. Whilst the calibration tables themselves contain both delay and amplitude calibration measures, the stability of the correlators meant that the primary calibrator was enough to ensure that flux densities were as correct as could be expected from the relatively short observation time of each source. For a typical source in the interior regions of a map that has no flagged data, there will be 6 observations of 54\,ms each --- three in a northward scanning interleave and three in a southward scanning interleave. This gives a total integration time of 324\,ms and a system equivalent flux density (SEFD) of $300-400$\,Jy, corresponding to an rms noise level  of $10-14$\,mJy/beam.

In the time between the scanning survey observations and the follow-up confirmation observations the absolute flux density scale for a particular regions was not accurately known. However, the stability of the WBAC meant that the relative flux density scale within a declination band was constant. When scheduling the follow-up observations, the strongest sources were given higher preference so that, in the absence of clustering, they were observed before the weaker sources. Once the follow-up observations were complete and the data had been reduced it was then possible to scale the scanning survey flux densities to that of the follow-up observations. This re-scaling does not add any extra information to any of the sources within the follow-up catalog, but it is useful when computing the completeness of the two surveys as discussed in section \ref{sec:completeness}.

\begin{figure}[hbt]
 \centering
 \epsfig{file=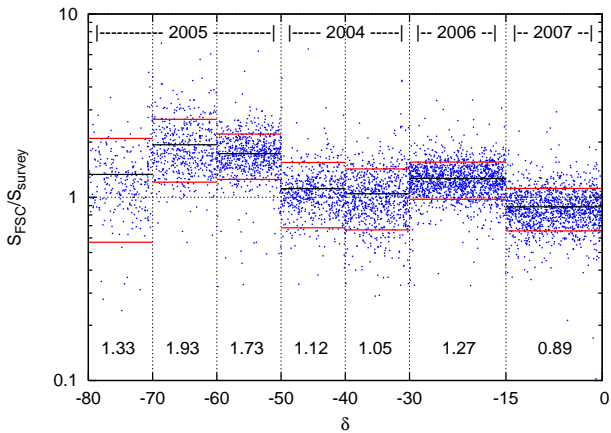,height=0.7\linewidth, angle=-90}
 \caption[Ratio of follow-up survey catalog and scanning survey fluxes]{Ratio of follow-up survey catalog and scanning survey fluxes. The horizontal black lines represent the mean flux ratio for each declination band, the values of which are printed below the data. The red lines represent a 1$\sigma$ deviation from the mean. The mean flux ratios were used to bring the survey map flux density scale in line with the more reliable follow-up observations. The year of scanning observations is given at the top of each declination band.}
 \label{fig:fluxscale}
\end{figure}

Figure \ref{fig:fluxscale} shows a comparison between the follow-up survey catalog (FSC) flux densities and the scanning survey flux densities as a function of declination. The scatter in the flux ratio is due to the fact that the follow-up observations were much more sensitive and have better $(u,v)$ coverage, allowing for a more accurate measure of the flux densities. The variation between regions that were observed contiguously is much less than that of regions that were observed in different years. The flux of the band-pass calibrator sets the flux density scale for each particular region. The flux of 1921$-$293 varies slowly enough that it possible to interpolate the observations of the AT calibrator monitoring project to obtain an accurate flux.

\section{Motivation for a Scanning Survey Catalog}
\label{sec:catalog}
Initially the purpose of the scanning observations was to detect individual sources so that more accurate, follow-up observations could be made of the stronger sources. The amount of time between scanning observations and follow-up observations was typically only 2-3 weeks in order to minimize the effects of source variability. This required that the calibration and production of candidate source lists be done as quickly as possible. The lists of candidate sources obtained from the scanning data were used only as an intermediary between scanning and follow-up observations. The fluxes and positions had low accuracy and early on in the survey there were many spurious candidate sources which decreased the efficiency of the follow-up observations. If the accuracy of the positions and fluxes of the candidate source lists could be improved, and the number of false positive sources reduced or at least well characterised, then the candidate source lists become a useful catalog in their own right. Once the AT20G follow-up observations were completed the calibration and map making process was able to be finalized and applied to each of the declination bands to produce more reliable candidate source lists. It also became possible to use the final AT20G (follow-up) survey catalog to gauge the effectiveness of different calibration and map making procedures. 

The remainder of this section is devoted to creating the best possible catalog of sources using the scanning survey data from the AT20G. To avoid confusion, the sources that constitute the final release version of the AT20G catalog \citep{murphy_australia_2010} will be referred to as {\em follow-up sources}, or as belonging to the {\em follow-up survey catalog} (FSC). The sources that were identified within the survey maps will be referred to as {\em scanning survey sources} or as coming from the {\em scanning survey catalog} (SSC). Although the FSC should logically be a sub-set of the SSC, this is not strictly true. Of the sources within the FSC 89\% of them are also within the SSC. The 11\% of sources are missing from SSC due to the flux and reliability cut off that was imposed -- most of the missing sources have a signal to noise less than 5, while the stronger missing sources have a reliability which is $<50\%$. Further details and discussion given in section \ref{sec:sourcedetection}.

A catalog produced from the scanning observations is able to cover a larger area of sky than the subsequent follow-up observations and to a lower limiting flux density. The reliability of the claimed sources within the scanning survey catalog will not be as high as if follow-up observations had been done, as there is no way to directly confirm a source's existence. A catalog which is more complete and yet less reliable than the follow-up survey catalog will be of particular use for the construction of foreground masks for CMB anisotropy missions. In the measurement of the CMB anisotropies, excluding a real source from the mask is far more detrimental than the inclusion of a noise spike, and thus the most complete catalog is the most useful, even when not all of the claimed sources are real. A secondary use of the scanning survey catalog is to identify sources that were excluded from the FSC due to observing time constraints or poor weather. If the reliability and completeness of the scanning survey catalog is characterised well enough, it is then possible to estimate the completeness of the FSC. As was reported by \citet{murphy_australia_2010}, the FSC is 91\% complete at 100\,mJy and 79\% complete at 50\,mJy.

\section{Imaging the sky}
\label{sec:imaging}
Once a suitable calibration scheme had been developed as described in section \ref{sec:calibration}, it was then possible to combine the scanning interleaves to produce a map of the sky. Due to the non-uniform spacing of the delays within the correlator it was not possible to create a map using standard Fourier inversion techniques as an accurate inverse transform is no longer guaranteed, and correlated noise signals will result. 

As each part of the sky falls within the primary beam of more than one observation (see figure \ref{fig:interleave}) it was necessary to create an auxiliary map of weights that would allow the maps to be normalised once all of the observations had been processed. Pixels were weighted according to their distance from the beam center and in proportion to the square of the primary power gain, and the weights were scaled so that a pixel that was observed at the beam center had a value of 1. By scaling the weights in this manner it was possible to use this auxiliary map to determine how well each part of the sky had been covered by the scanning observations. This sky coverage map was used to give quantitative confidence levels in the source detection stage (section \ref{sec:sourcedetection}), as well as serving as a way to determine the survey sky coverage (section \ref{sec:coverage}).

\begin{figure}[hbt]
 \centering
 \epsfig{file=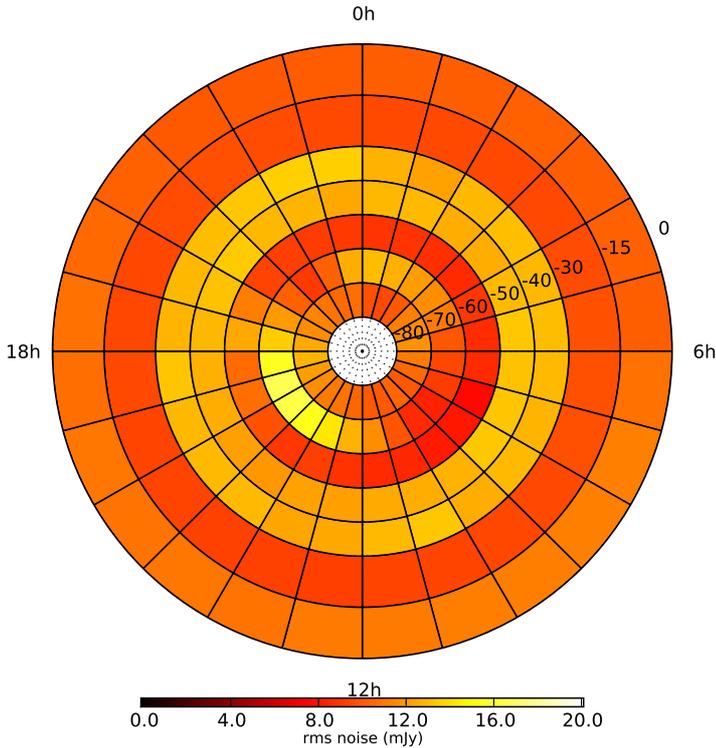,bb=70 135 540 630, width=0.8\linewidth,clip=}
 \caption[The corrected rms noise level for the survey maps]{The corrected rms noise level for the survey maps. The rms noise level is consistent with an SEFD of 300-400\,Jy. Variations within a declination band are due to variations in the survey coverage.}
 \label{fig:rms_noise}
\end{figure}

The SEFD, sky coverage and calibration effectiveness are all critical to producing maps with high signal to noise. Figure \ref{fig:rms_noise} shows the rms noise level for each of the maps within the survey region. The typical rms noise is around 10\,mJy/beam, which is in agreement with that calculated in section \ref{sec:fluxcalibration} for a SEFD of 300\,Jy. The region $-30\deg\ge\delta\ge-50\deg$ has a higher rms noise value of around 14\,mJy which is consistent with the measured SEFD of 400\,Jy. The variation in rms noise level seen in the region $-60\deg\ge\delta\ge-70\deg$ around $\alpha=15^h$ is due to patching observations that were done using the incorrect interleave number and thus areas with poor data were not recovered.

The resulting (flux density) map of the sky remains convolved with the telescope response pattern. A CLEAN algorithm was not able to be used to deconvolve the map as a Fourier inversion would not produce a map consistent with our data. A specialised source detection program was created that would match sources to the telescope's beam pattern as measured by the primary calibrator. Further details are discussed in section \ref{sec:sourcedetection}. 

\section{Sky Coverage}
\label{sec:coverage}
\begin{figure}[hbt]
  \centering
  \epsfig{file=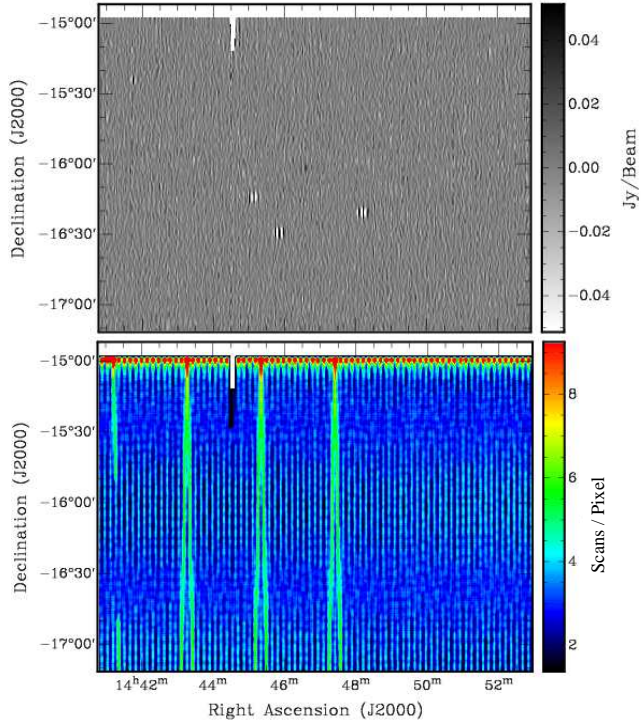, width=0.7\linewidth,clip=}
  \caption[Convolved sky map and sky coverage for a small region of the sky]{{\bf Top:} Convolved sky map for a small region of the sky. There is a small amount of sky missing at the top of the map due to flagged data. Three sources with flux densities of 222\,mJy (left), 286\,mJy (middle), and 284\,mJy (right) are easily visible. {\bf Bottom:} Sky coverage map for the same region. Areas in white have no data, whilst black areas have a coverage lower than 2 scans/pixel. The effect of flagged data on the northern boundary can be seen. The scan in green is an instance of patching data that overlaps with good data and thus increases the sky coverage for a small part of the sky.}
  \label{fig:sky_maps}
\end{figure}

As mentioned in section \ref{sec:imaging} the procedure that created the sky maps also created a map of the pixel weights which, when correctly scaled, become a map of the sky coverage. Figure \ref{fig:sky_maps} shows a region of sky and the corresponding coverage map. The map of the sky coverage reveals a lot about how the survey was conducted and what sorts of problems were encountered. The interior parts of the map are scanned at least twice with a separation between scans being of the order of days, and the scans are in opposite directions. Figure \ref{fig:sky_maps} shows that the typical sky coverage is around $3\pm0.5$ scans/pixel, with variation due to the way that the different scans overlap at each point on the sky. As the telescopes require some time to turn around at the northern and southern boundaries of each declination band, the spacing between observations decreases and the sky coverage increases dramatically toward the boundaries. The coverage at the turn-around points of the scans can be as much as 20 scans/pixel. When data are flagged for poor quality on the interior of the map, and no patching has been done to recover the data, there is a reduction in the sky coverage at that point. The boundary regions of the maps, whilst still having a twofold redundancy in coverage, have a scan separation of only a few seconds in time, and thus data flagged for a contiguous block of time removes scans in both directions. The small gap in the map at $14^h44^m20^s -14\deg57'23"$ is a result of data that were flagged out due to the presence of strong radio emission from Jupiter. When a significant amount of data are lost due to bad weather or otherwise, an attempt is made to re-observe the data and the patching data generally overlaps the flagged data by a minute or more. The result in this over-patching is better coverage for some parts of the sky. This is also evident in figure \ref{fig:sky_maps} where the coverage increases to almost 6 scans/pixel in some places.

For the most part, variations in the sky coverage do not affect the quality of the scanning survey catalog as they are only small. However there were problems with the zenith ($-30\deg18'46''$ for the ATCA), that caused a lot of data to be flagged at the hardware level and were not able to be recovered. The large amount of flagged data around $-32\deg$ declination causes the sky coverage to drop below 1 scan/pixel. The amount of sky that is affected is small but the areas of low coverage produce many spurious sources that make the detection of real sources in this area quite difficult and thus decreases the completeness of the survey.

\section{Source Detection}
\label{sec:sourcedetection}
The way in which candidate sources were detected using the scan data changed over the course of the AT20G observations. In 2004 the scan data were processed as time ordered data using a stream-based approach. The final candidate source detection of 2007 was based on the scan maps discussed in the previous section. Intermediate years used a combined approach with the detection rate of the followup observations increasing year by year. In each case it was necessary to have a fast turn around for the candidate source lists, as follow-up observations were typically scheduled within 2-3 weeks of scanning observations to reduce the effects of variability. The two detection methods are discussed below.

\subsection{Stream-Based Detection}
In the early years of the survey, before the intra-scan calibration methods had been refined, candidate source lists were required quickly, so that follow-up observations could be done. The following equation:
\begin{equation}
S(\tau_{ast}) = \sum {C(\tau,\tau_{ast}) C_{obs}(\tau)}
\label{eq:selfcal}
\end{equation}
where $C_{obs}$ are the measured correlator data, and $C(\tau,\tau_{ast})$ system response to a calibrator (Figure \ref{fig:lag_response}), was used to identify candidate sources whenever a signal to noise of 5 or more was reached. This method was relatively fast and easy to compute but was less sensitive to sources that were towards the half power point of the primary beam. Also, due to the lack of intra-scan calibration, sources that appeared in two separate data streams often had position and flux discrepancies. The problem of side-lobe and multiple detections was significant and many side-lobe or duplicate sources entered into the follow-up observations. Once a follow-up observation had been made such occurrences were easily detected. In 2004 the confirmation rate of the follow-up survey was as low as 50\%, meaning that up to half of the observing time allocated to follow-up observations resulted in non-detections. In 2005 the detection scheme, calibration, and duplicate/side lobe rejection methods were improved. The problem of reduced sensitivity to sources that were at positions not scanned close to the beam center remained. During 2005 the calibration scheme was improved to such an extent that a map based detection method was comparable in reliability to the stream based method, and a correlation of the two results was used as a candidate source list. From 2006 onwards the map based detection method was reliable enough that it was used alone.

\subsection{Map-Based Detection}
\begin{figure}[hbt]
 \centering
 \epsfig{file=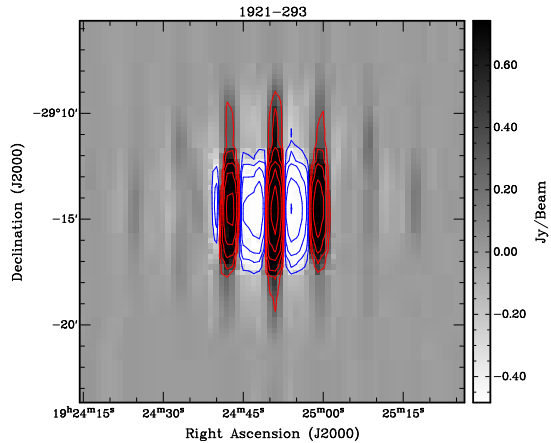, height=0.6\linewidth, angle=-90, clip=}
\caption[Calibrator template]{The convolved image of primary calibrator 1921$-$293, which was used to create the template used in the detection of sources within the convolved sky maps. The lowest contours are (red/blue) $\pm 0.25$\,Jy/beam with consecutive contours doubling up to $8$\,Jy/beam. Primary beam side-lobes can be seen outside of the contours at a few percent of the source flux of 11\,Jy.}
\label{fig:ps_response}
\end{figure}

As the final maps were still convolved with the antenna response, it was necessary to create a specialised source detection program (detailed in an upcoming ATNF memo). The program works in a very CLEAN-like manner: searching for the largest peaks in the map, fitting to a predetermined model, and removing it from the map. The primary calibrator was used to determine the antenna response which makes up the template against which candidate sources are matched. The antenna response to the primary calibrator PKS\,1921$-$293 is shown in figure \ref{fig:ps_response}. Starting at the highest peak in the map an iterative process was run where by the template was matched to the peak, a flux and position were recorded along with: a measure of how well the source response matched the calibrator template, the number of pixels that were within the fitting area, the number of blanked pixels within the same area, and the sky coverage at the source position. A region of the map around the peak was then blanked and the process continued until some minimum flux density was reached. The maps were produced in blocks of $1^h6^m$ in RA and between $10\deg$ and $15\deg$ in DEC. The extra six minutes of overlap in RA between the map and its neighbours avoids problems involved when detecting a source on the boundary of a map.

The searching algorithm does not take account of possible extended sources. For the 30\,m survey spacing there is a 50\% reduction in amplitude for a Gaussian source with size 45\,arcsec, and negligible response for sources much larger than 1.5\,arcmin. Any sources that are extended will have underestimated flux density in these survey maps. Fornax A, a well known extended radio source at low frequencies, is present within the maps at a level of 50\,mJy. All of the extended emission is completely resolved out. A point source that falls within a beam width of a stronger source will not be detected in the source detection program as there is no allowance given for double or multiple sources, however such sources will become evident during follow-up observations. The follow-up observations have an increased $(u,v)$ coverage and integration time, resulting in a better position and flux density measurement. The FSC does not suffer as badly from the problem of missing flux for extended sources, although a source with no bright core or hot spots will likely not be detected in the survey maps and thus not make it into the follow-up observing schedule and thus not be present in the FSC.

The list of candidate sources found in the maps was filtered to remove side-lobes of strong sources and sources that were detected twice due to their position at the edge of two different maps. The Galactic plane is a plentiful source of extended and close multiple objects of sufficient strength to confuse the source detection algorithm and are not well characterised in the maps. Candidate sources with a Galactic latitude of $|b| \le 1.5\deg$ were thus removed from the list. Areas of the sky which are close to the declination boundaries, or which have flagged or missing data in one or more interleaves, will have a much higher noise than surrounding areas. Since the source detection algorithm uses only a single noise value for an entire 1\,hour by $10\deg-15\deg$ map these regions will increase the overall noise level by only a small amount, but contribute many false sources with a flux more than five times the map rms noise (see Figure \ref{fig:lfmatching}). Spurious sources were identified and removed via a filtering metric which is now discussed.

\subsection{Source Filtering}
In order to increase the reliability of the scanning survey catalog, a filter was used to remove sources that were likely to be spurious.. Lower frequency catalogs were used to guide the creation of the filter but were not in the actual source selection. Once the initial candidate list had been prepared it was matched against lower frequency sky surveys to determine what limits should be placed on the filtering process to minimise both the number of spurious sources included and real sources excluded. A combination of the Sydney University Molonglo Sky Survey (SUMSS version 2.1 of March 11 2008, \citet{mauch_sumss:wide-field_2003}), and the Second epoch Molonglo Galactic Plane Survey (MGPS-2 version 1.0 of August 15 2007, \citet{murphy_second_2007}) was used to identify sources within the candidate source lists that were likely to be real sources. In regions above a declination of -30\deg the NRAO VLA Sky Survey (NVSS, \citet{condon_nrao_1998}) catalog was used. Whilst it was not expected that the lower frequency catalog would be completely recovered, it was known from previous observations (in particular \citet{ricci_first_2004}), that almost all sources detected at 20\,GHz were previously known in the more sensitive lower frequency surveys. 

A source was considered to be identified if had a low frequency counterpart within 1 arcmin. Some follow-up survey catalog (FSC) sources were initially found to have no counterpart in the relevant catalog, but closer inspection of the survey maps revealed extended sources that were either not listed in the scanning survey catalog (the MGPS-2 catalog contains only the point sources), or that have a hot-spot that is some distance from the FSC position. Of the approximately 6000 sources within the AT20G FSC only 25 were unidentified \citep{murphy_australia_2010}. This result is consistent with that of \citet{ricci_first_2004}, who found low frequency counterparts to all 221 sources detected at 18GHz.

\begin{figure}[hbt]
 \centering
 \epsfig{file=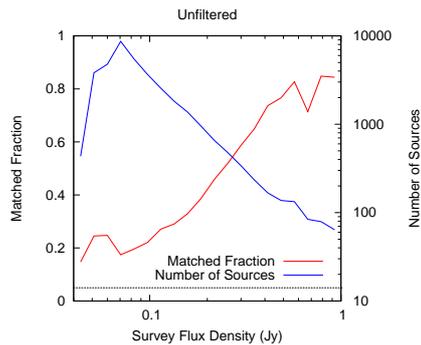,bb=210 55 555 500, height=0.49\linewidth,angle=-90,clip=}
 
 \caption{Low frequency matching rate as a function of corrected survey flux at 20\,GHz for the unfiltered scanning survey catalog. The blue line shows the number of sources per flux bin, whilst the red line shows the fraction of sources matched per flux bin. The dotted line represents the random matching probability of 5\%.}
 \label{fig:lfidrate}
\end{figure}

The matching rate between the scanning survey catalog and the low frequency catalogs is shown in figure \ref{fig:lfidrate}, where a strong correlation between flux density and matching rate can be seen. This correlation is expected since the number of spurious sources in the scanning survey catalog increases as the survey flux decreases. The anti-correlation with source count is expected for the same reason. 

\begin{figure}[hbt]
\centering
 \epsfig{file=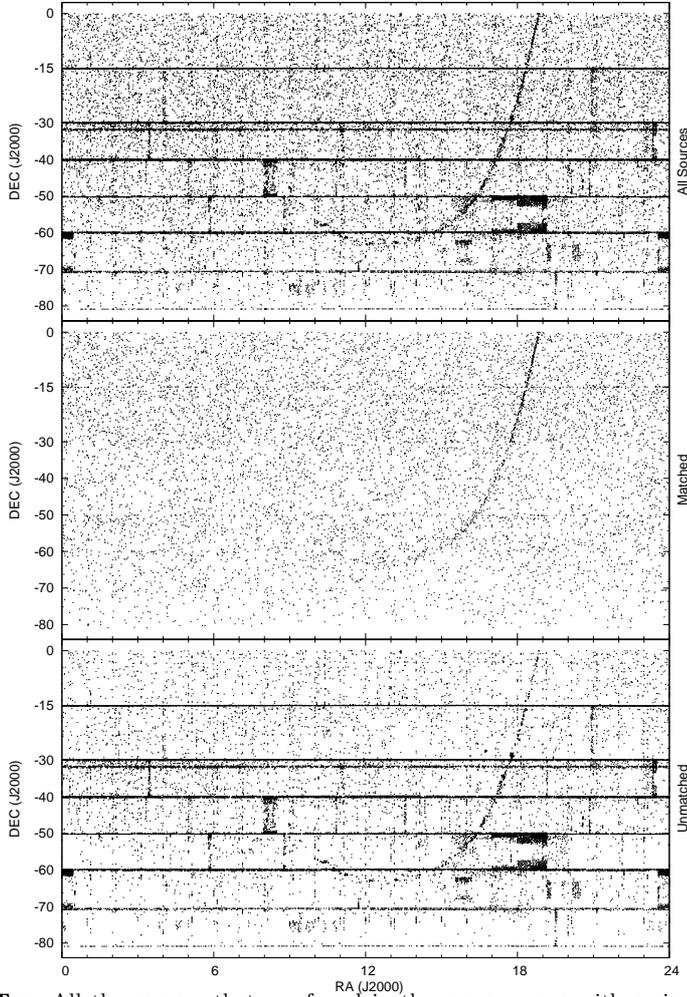, height=0.77\linewidth,angle=-90}
\caption[All the sources that are found in the survey maps]{{\bf Top}: All the sources that are found in the survey maps with a signal to noise greater than 5. {\bf Middle}: Sources that are matched with a low-frequency source. {\bf Bottom}: Sources that are not matched with a low-frequency source. The matching with  the NVSS, SUMSS, and MGPS-2 catalogs is a good indicator of sources that are likely to be real.} \label{fig:lfmatching}
\end{figure}

Figure \ref{fig:lfmatching} shows all of the sources that were detected in the survey maps with an SNR $>5$. The raw scanning survey catalog shows features that can clearly be attributed to noise, such as the over-density of sources towards the boundary of each declination zone. Variations in the survey coverage are also a cause of spurious sources and can be seen around 18\,hours in the $-50\deg\ge\delta\ge-60\deg$ zone. The zone $-30\deg\ge\delta\ge-40\deg$ contains the zenith as seen by the ATCA and thus has a region of poor coverage interior to the map. The overlapping of sources towards the zone boundaries is seen at a much lower level in the matched source plot, which can be attributed to sources that were detected in the overlapping regions of maps from different declination zones. It can be seen from Figure \ref{fig:lfmatching} that matching with low frequency catalogs is an effective way to distinguish between real and spurious sources. The correlation with lower frequency surveys was used as a guide to determine what combination of the quality control quantities would produce the largest fraction of confirmed detections. The low frequency surveys were used to determine what a real source would look like within the map; a source's presence in a lower frequency survey was not part of the filter, so as not to bias the final catalog. At sufficiently low 20\,GHz flux densities the correlation between low frequency detection and followup confirmation is expected to break down. For a source to be detected at 50\,mJy in the AT20G a spectral index of $\alpha^{20}_{1.4,0.8}\geq +1.3$ would be required to avoid detection in the NVSS, SUMSS, or MGPS-2 surveys. This rare class of ultra inverted spectrum sources is discussed in \citet{massardi_australia_2008,murphy_australia_2010}.

With a matching radius of 60\,arcsec, some matches will be due to chance and not indicative of any real identification between the two sources. \citet{condon_nrao_1998} give a probability of 5\% for finding an unrelated radio source within 60\,arcsec of an NVSS source. The SUMSS and MGPS-2 catalogs have source densities of 30 sources per square degree, whilst the NVSS has 50 sources per square degree. Thus we expect that around 5\% of the identifications within the middle plot of figure \ref{fig:lfmatching} are spurious. Figure \ref{fig:lfidrate} shows a minimum matching rate that is not zero and can be attributed to this effect.

To distinguish between real sources and those that are likely to be due to poor coverage or high noise peaks, a filtering metric was established with the guidance of a genetic algorithm. It was found that the metrics that produced the greatest separation between the matched and unmatched source populations were variations on equation \ref{eq:filter}. Here $S_{20}$ is the flux in Jy as measured in the survey map, $\chi^2$ is the cross correlation of the source and the calibrator template, cov is the coverage as measured in the auxiliary map, and pix is the ratio of good to bad pixels that were within the blanking region around the source.

\begin{equation}
\mathcal{F} = \log_{10}\left(S_{20}^2 . \chi^2 . \mathrm{cov} . \mathrm{pix}\right)
\label{eq:filter}
\end{equation}

\begin{figure}[hbt]
 \centering
 \psfrag{curlyF}{$\mathcal{F}$}
 Unmatched at low frequency \hspace{0.2\linewidth} Matched at low frequency
 \newline
 \epsfig{file=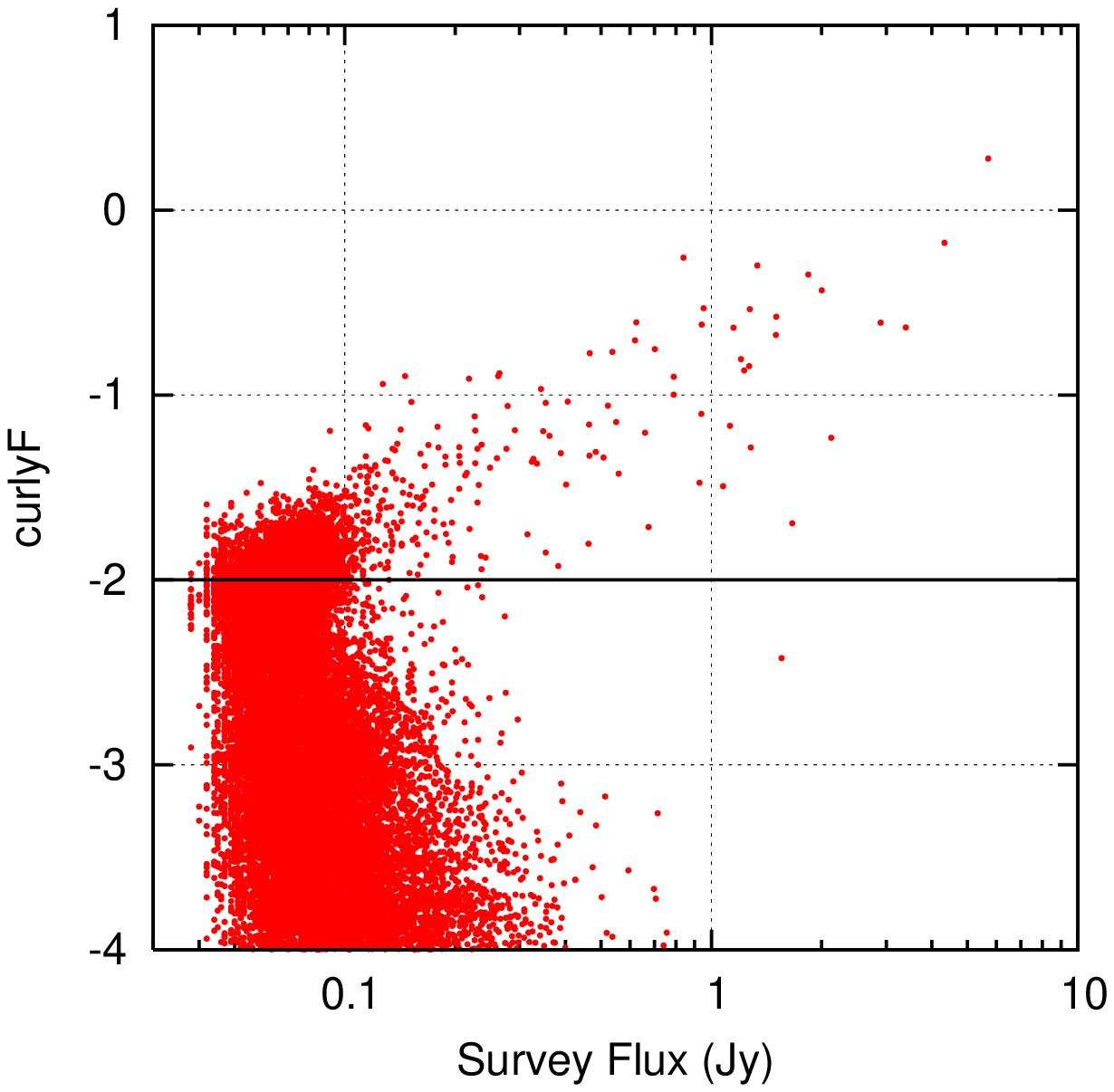,bb= 210 55 555 410, height=0.45\linewidth, angle=-90}
 \epsfig{file=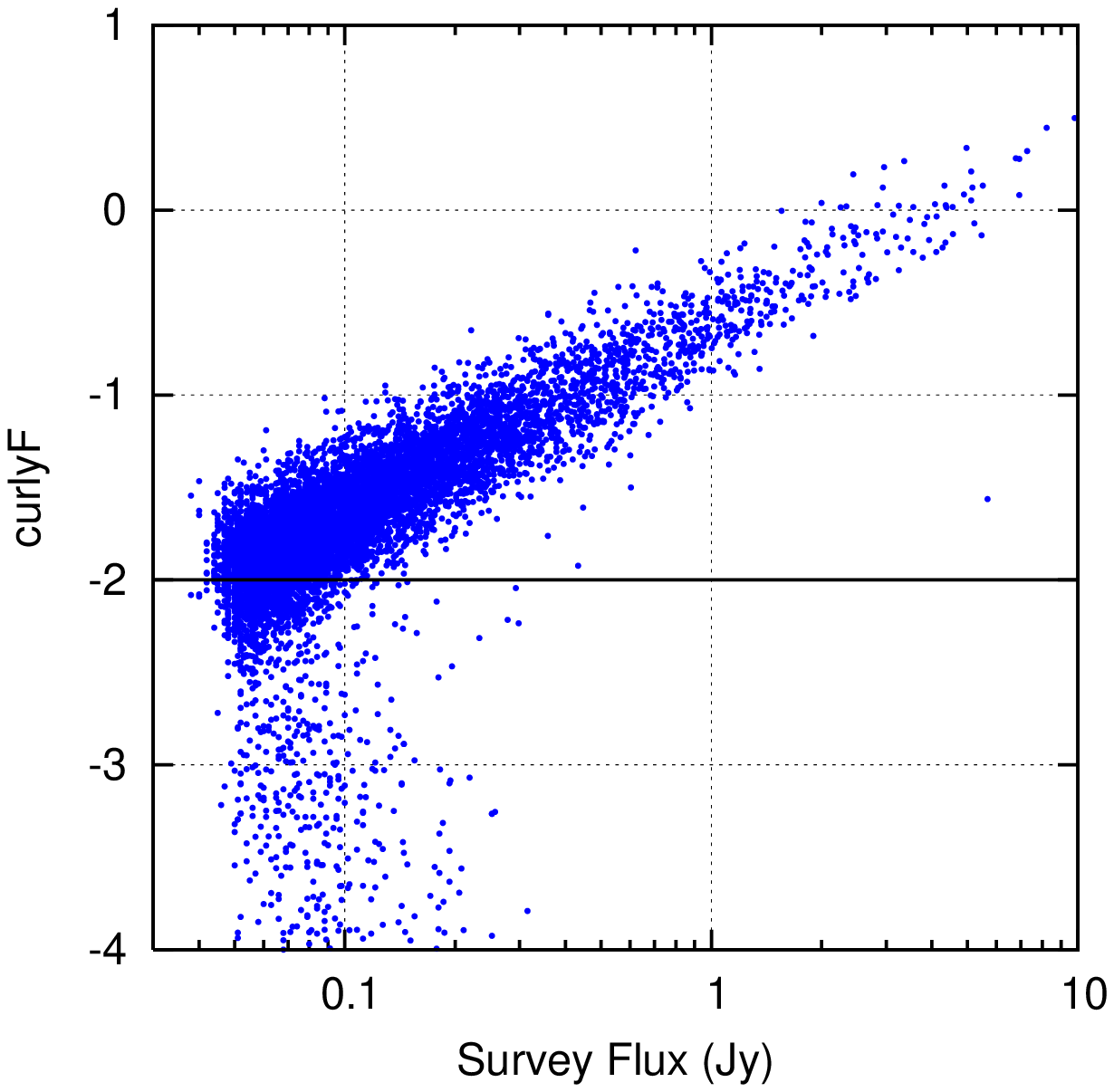  ,bb= 210 55 555 410, height=0.45\linewidth, angle=-90}
 \caption[$\mathcal{F}$ as a function of survey flux, matched and unmatched at low frequency]{The filtering metric $\mathcal{F}$ as a function of survey flux for the scanning survey catalog. Blue sources ({\bf right}) are those that have a low frequency counterpart within a 1 arc-min radius, red sources ({\bf left}) are those that do not. The horizontal line represents the point at which equal numbers of sources are matched and unmatched.}
 \label{fig:filterlf}
\end{figure}

\begin{figure}[hbt]
 \centering
 \psfrag{curlyF}{$\mathcal{F}$}
 Unmatched with FSC \hspace{0.2\linewidth} Matched with FSC 
 \newline
 \epsfig{file=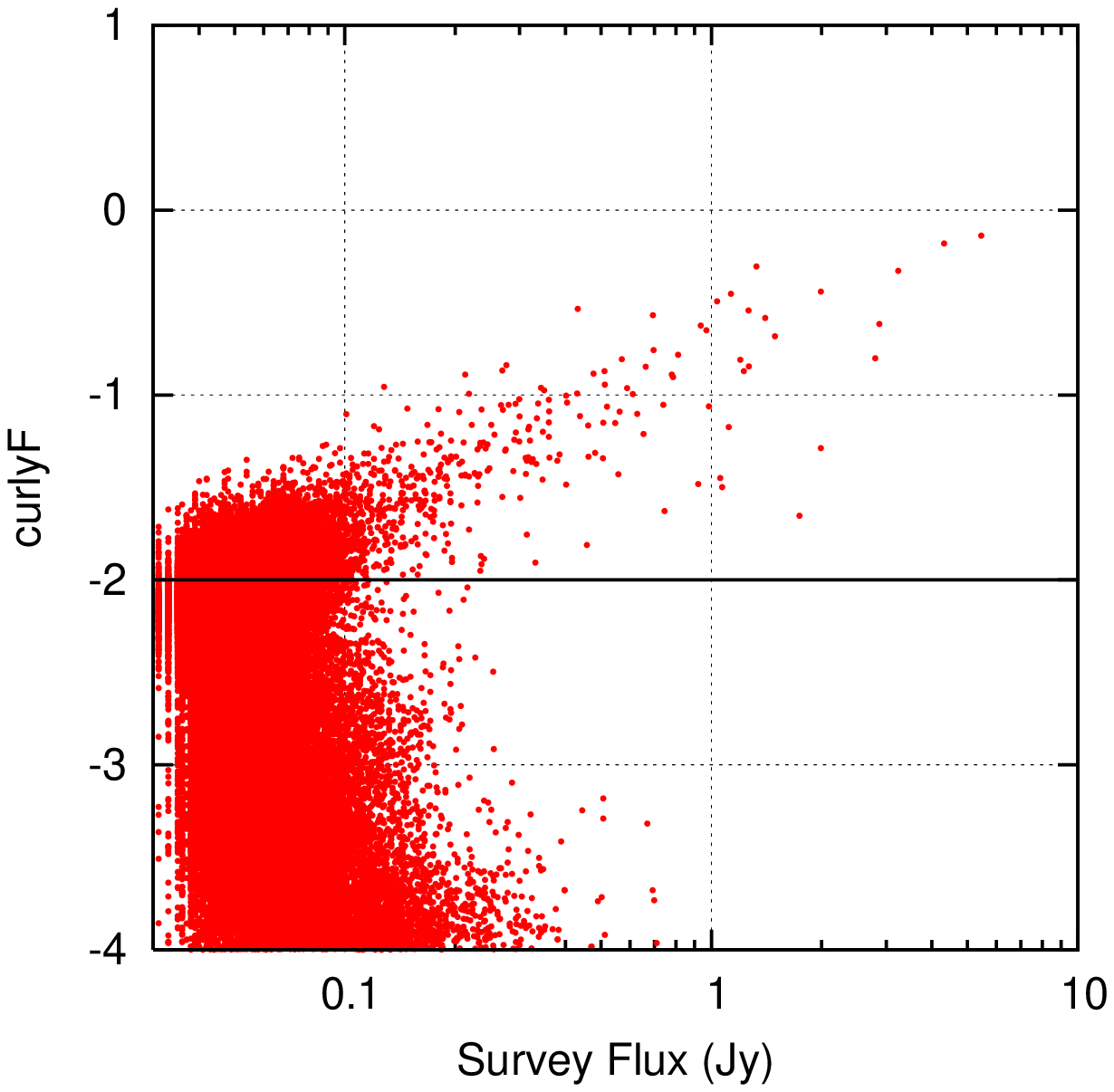,bb= 210 55 555 410, height=0.45\linewidth, angle=-90}
 \epsfig{file=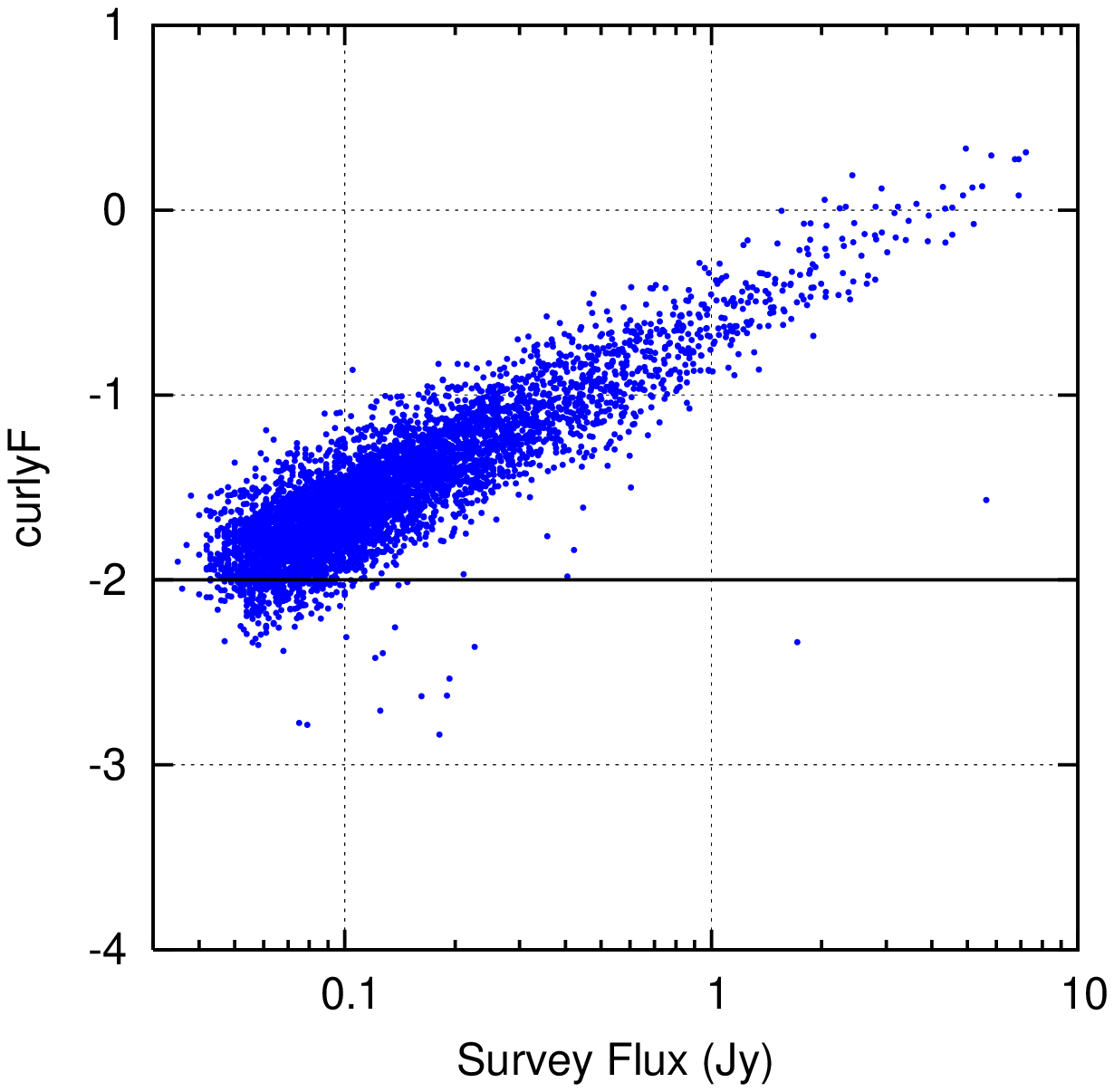,  bb= 210 55 555 410, height=0.45\linewidth, angle=-90}
 \caption[$\mathcal{F}$ as a function of survey flux, matched and unmatched at 20\,GHz]{The filtering metric $\mathcal{F}$ as a function of survey flux for the scanning survey catalog. Blue sources ({\bf right}) are those that have a follow-up survey catalog counterpart within a 1 arc-min radius, red sources ({\bf left}) are those that do not. The horizontal line represents the point at which equal numbers of sources are matched and unmatched at low frequency.}
 \label{fig:filterfsc}
\end{figure}

Figure \ref{fig:filterlf} shows $\mathcal{F}$ as a function of survey flux for sources within the scanning survey catalog, identified via matching with the low frequency catalogs.  The horizontal line at -2 represents the point above which there are an equal number of matched and unmatched sources. A catalog constructed with only sources that have $\mathcal{F}\ge -2$ will have an overall matching rate of 50\%. It is expected that nearly all the real sources at 20\,GHz will have a low frequency counterpart, however if that counterpart is complex or extended then a more careful match between the catalogs will be required to identify counterparts. For this reason it is expected that some of the 20\,GHz sources with $\mathcal{F}\ge-2$ will remain unmatched despite the fact that they are real sources. Similarly not all the sources that are matched at low frequency are expected to be real. In figure \ref{fig:filterlf}({\em right}) there are 1758 sources below $\mathcal{F}=-2$, which is 4.7\% of the 35598 unmatched sources in the same area of the plot. This is in agreement with the 5\% random matching rate cited for the low frequency surveys previously. The matched and unmatched sources occupy the different, yet overlapping, areas of ($\mathcal{F}$,S) space. Using the low frequency identification as a proxy for the detectability of a survey source, it is apparent that eq \ref{eq:filter} is efficient at separating real sources from noise peaks. Figure \ref{fig:filterfsc} is the same as \ref{fig:filterlf} with the sources matched with the FSC. The same separation of sources can be seen here, indicating that the use of the low frequency catalogs is a good proxy for source detectability. 

\section{Reliability}
\label{sec:reliability}
The presence of an apparent source within the survey map does not guarantee that it is a real source that will be confirmed in a followup observation. The rate at which false positive detections are made by the source finding program should be taken into account.

\begin{figure}[hbt]
 \centering
 \psfrag{curlyF}{$\mathcal{F}$}
 \psfrag{curlyR}{$\mathcal{R}$ \%}
 \epsfig{file=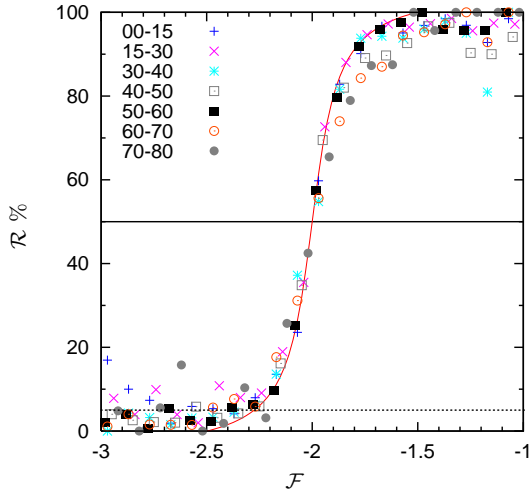, height=0.8\linewidth,angle=-90}
 \caption[The fraction of sources that were matched with a low frequency catalog, as a function of the filtering metric]{The fraction of sources that were matched with a low frequency catalog, as a function of the filtering metric. The red curve represents the relation between $\mathcal{R}$ and $\mathcal{F}$ that is used to assign a reliability estimate to individual sources. The lower horizontal dotted line represents the spurious matching rate of 5\%, whilst the upper solid line is the point at which half of the sources are matched. }
 \label{fig:reliability}
\end{figure}

To estimate the reliability for each region of the sky within the AT20G a comparison was made between the sources that were found in the source detection process and a low frequency catalog that covered the same area of sky. Figure \ref{fig:reliability} shows how the low frequency matching rate changes with the filtering metric. The rate at which sources are found within one of the three low frequency catalogs is used as the reliability and is denoted by $\mathcal{R}$. At $\mathcal{F}$ values below -2.5 the fit approaches 0\% and above -1.5 it reaches 100\%. Neither of these two values are considered to be practical and the reported reliability, $\mathcal{R}$, of a source is limited to $5\% \leq \mathcal{R} \leq 98\%$.

Once the relationship between $\mathcal{F}$ and $\mathcal{R}$ had been established it was then possible to assign a reliability estimate to each source within the scanning survey catalog. Sources that were assigned a reliability of $\mathcal{R}<50\%$ were removed from the final version of the scanning survey catalog.

\section{Survey Completeness}
\label{sec:completeness}
In theory a comparison between a catalog and a list of all real objects gives an accurate measure of the catalog completeness, however an {\em a priori} list of all real objects does not exist.

Due to the many calibration steps that were performed at the software level rather than the hardware level, and the relatively static performance of the correlator itself, it was possible to know a lot about how the system responded to a given source. Because of this it was possible to insert artificial sources into the data stream at the most basic level and then follow them through the calibration, map making, source finding and filtering process. Having this amount of control over the calibration and source detection process meant that it was possible to create an accurate measure of the scanning survey catalog completeness, and by extension, the follow-up survey catalog completeness.

Survey completeness was estimated in a semi-analytical manner. In order to obtain a set of data that were consistent with the actual observations, whilst including as many of the detrimental factors as possible it was decided that the survey data would be used. Specifically, the regions between the detected sources were considered to be an excellent resource for the compilation of a data set that could be used to estimate the completeness and reliability of the survey.

The injection of sources was achieved by inserting a scaled copy of the primary calibrator into the data stream as it was being processed. This method is sensitive to relative timing errors (ie. between scans) but not to absolute timing errors. The absolute timing errors are detected via comparison of the final maps to known calibrators or the follow-up survey catalog positions and fluxes. The completeness estimate is thus assuming that the timing calibration is correct to first order. 

For each $1^h\times10-15\deg$ map of the sky, a list of sources was produced that had a specific range of fluxes and a random distribution of positions. The source positions were constrained so that they were not within a primary beam width of an already detected source. At 20\,GHz the density of sources detected is of the order $0.3\,\mathrm{deg}^{-2}$, and as the number of sources injected was around 500 per map, the injected source positions were constrained so that the measured completeness would not be affected by excess amounts of confusion. In this way it was possible to create a map of the sky that contained all of the properties of the real observations that would otherwise be hard to characterise and reproduce in a simulated map. Such properties include but are not limited to: sensitivity variations, regions that were subject to patching (possibly multiple times), T$_{\mathrm{sys}}$ variations due to weather that were not captured in the calibration process, and data flagging due to solar system objects\footnote{The Moon, Sun and Jupiter were troublesome for bands above $-30\deg$ declination}. It should be noted that the completeness measurement is made using only the injected sources and thus deriving a source count model for the scanning survey catalog using the completeness and reliability measures can be done in a non-circular manner.

\begin{figure}[hbt]
 \centering
 \psfrag{curlyR>00}{$\mathcal{R}\ge 0$}
 \psfrag{curlyR>90}{$\mathcal{R}\ge 90$}
 \epsfig{file=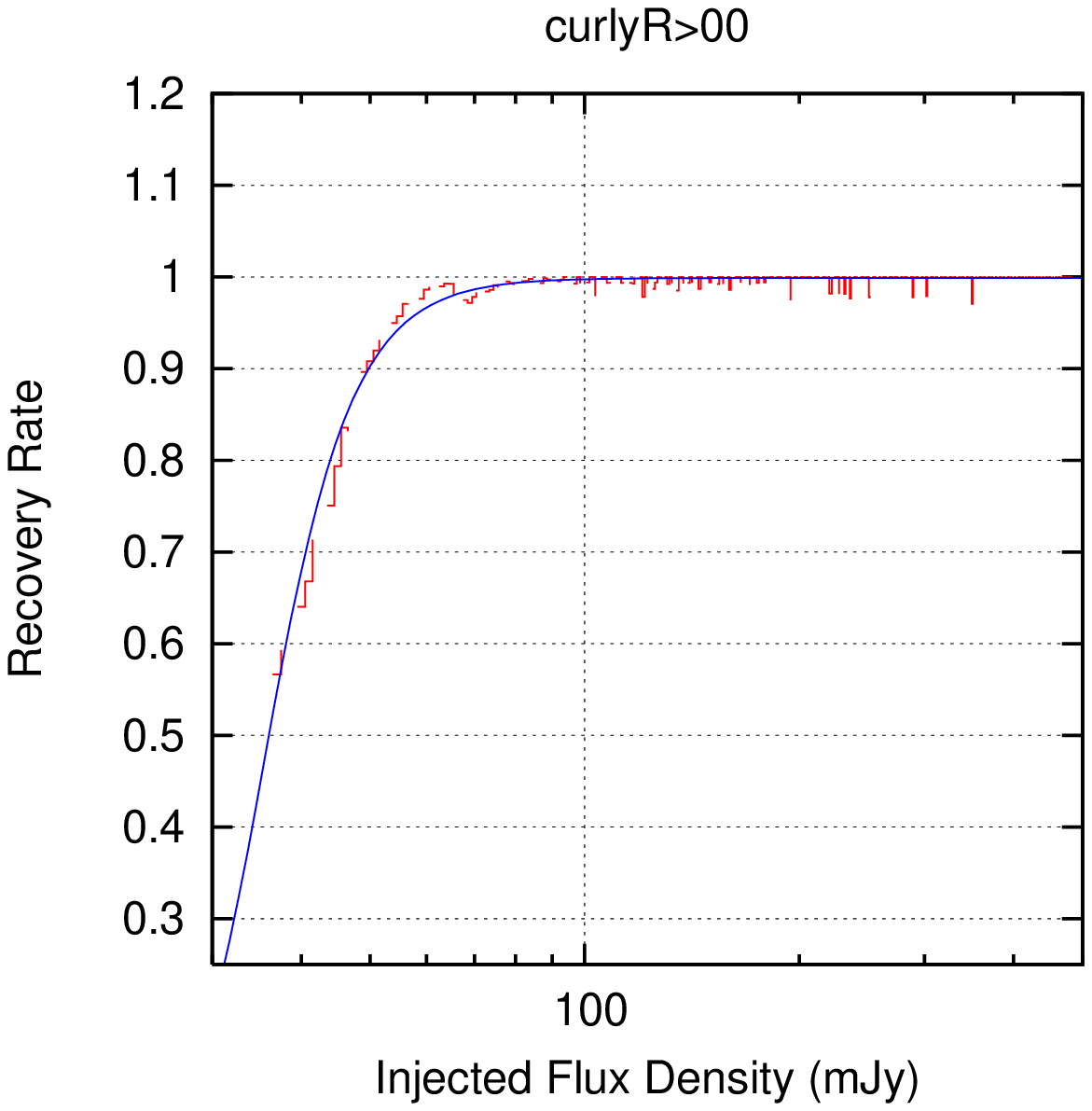,bb= 210 55 555 400, height=0.45\linewidth,angle=-90}
 \epsfig{file=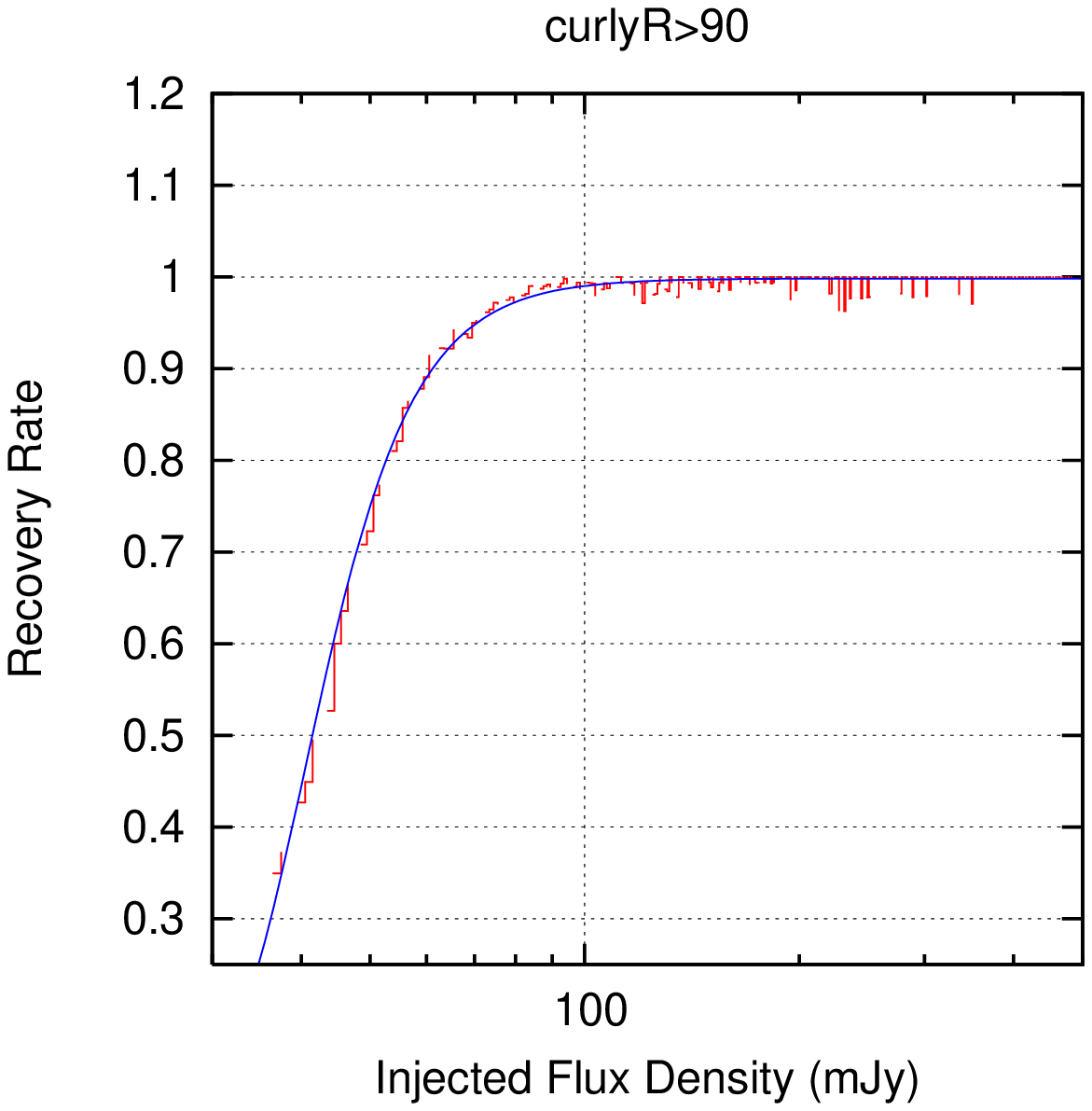,bb= 210 55 555 400, height=0.45\linewidth,angle=-90}
 \caption[The recovery rate for injected sources within the region $-15\deg\ge\delta\ge-30\deg$]{The recovery rate for injected sources within the region $-15\deg\ge\delta\ge-30\deg$. {\bf Left:} All sources. {\bf Right:} Sources with $\mathcal{R}\ge 90\%$. The red bars are the fraction of sources that were recovered from the maps, with the blue line being a fit to equation \ref{eq:recovery}.}
 \label{fig:recovery}
\end{figure}

The survey completeness was determined by first measuring the rate at which sources were recovered from the injected maps as a function of flux. Figure \ref{fig:recovery} shows the recovery rate for all maps within the region $-15\deg\ge\delta\ge-30\deg$. To eliminate the effects of noise within the recovery rate measurements, the function in equation \ref{eq:recovery} was fit to the data, where S is the flux, a is the maximum completeness and b,c are shape parameters.

\begin{equation}
 \mathcal{G}(S)= a\left(\frac{S}{b}\right)^c\left[1-e^{-\left(\frac{S}{b}\right)^{-c}}\right]
 \label{eq:recovery}
\end{equation}

Once a fit to equation \ref{eq:recovery} had been obtained for each of the maps, the recovery rate was converted to a source count using the source count model of \citet{sadler_extragalactic_2008} in equation \ref{eq:sourcecount} as shown in table \ref{tab:dndsparams}. The survey completeness at a flux S is then given by 
\begin{equation}
 \mathrm{C}(S) = \frac{\int_{S}^{S_{max}} \mathcal{G}(S) N(S)}{\int_{S}^{S_{max}}N(S)} 
 \label{eq:completeness}
\end{equation}
where $S_{max}$ is sufficiently large that the completeness is no longer changing. Although the source count model of \citet{sadler_extragalactic_2008} has uncertainties of $20-50\%$ in the actual number of sources within a particular flux bin, the completeness estimate outlined above is sensitive only to errors in the {\em relative } number of sources within each flux bin. Thus only the errors in the slope of the source counts ($\alpha$ and $\beta$) will affect the completeness estimates. The uncertainties in $\alpha$ and $\beta$ are 3\% or less, and when integrated over the range of 50\,mJy to 500\,mJy the total source count deviates by less than 9\%, but this affects the calculation of equation \ref{eq:completeness} by less than 1\%.

\begin{table}[hbt]
\centering
 \begin{tabular}{cc}
  \begin{minipage}[t]{0.45\linewidth}\centering
  \begin{tabular}{lc}
  \hline 
  \hline
parameter & value \\
\hline
 $\alpha$ & $1.92^{+0.03}_{-0.02}$ \\
 $\beta$  & $2.78^{+0.09}_{-0.07}$ \\
 $S_*$    & $1.09^{+0.13}_{-0.18}$ Jy\\
 $N_*$    & $28.9^{+13.7}_{-6.8} $ Jy sr$^{-1}$\\
\hline
 \end{tabular}
 \end{minipage}
&
\begin{minipage}[t]{0.45\linewidth}\centering
\vspace{-1.5cm}
\begin{equation}
N(S) = \frac{2N_*}{\left[ \left(\frac{S}{S_*}\right)^\alpha + \left(\frac{S}{S_*}\right)^\beta\right]}
\label{eq:sourcecount}
\end{equation}
\end{minipage}
\\
\end{tabular}
\caption{The source count model at 20\,GHz as reported by \citet{sadler_extragalactic_2008}.}
\label{tab:dndsparams}
\end{table}

\begin{figure}[hbt]
 \centering
 \epsfig{file=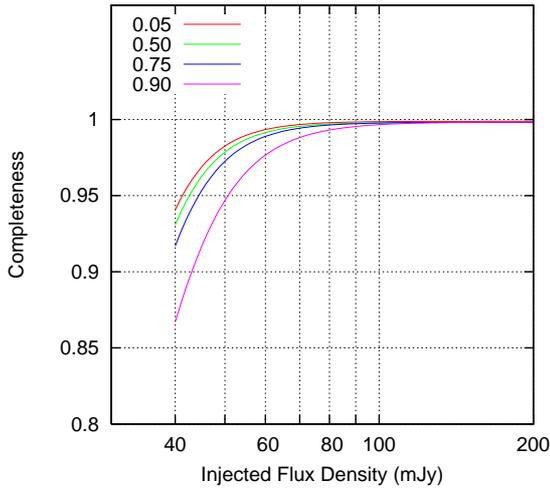, height=0.8\linewidth,angle=-90}
 \caption[The level of completeness as a function of corrected survey flux for the region $-15\deg\ge\delta\ge-30\deg$]{The level of completeness as a function of corrected survey flux density as measured by the injection and recovery of sources for the region $-15\deg\ge\delta\ge-30\deg$. The different coloured lines represent sources with a reliability estimate of $\mathcal{R}\ge (0.05,0.5,0.75,0.9)$. The completeness is given by equation \ref{eq:completeness}}
 \label{fig:completeness_plot}
\end{figure}

Figure \ref{fig:completeness_plot} shows the corresponding completeness levels for the plots of figure \ref{fig:recovery}. The effect of filtering is also demonstrated, and can be seen to lower the completeness of the catalog as the reliability is increased. Source that were within 1.5\deg of the Galactic plane are excluded from the follow-up survey catalog and are thus not included in the completeness calculations. The Galactic sources thus do not contaminate the completeness of the maps.

\begin{figure}[hbt]
 \centering
 \epsfig{file=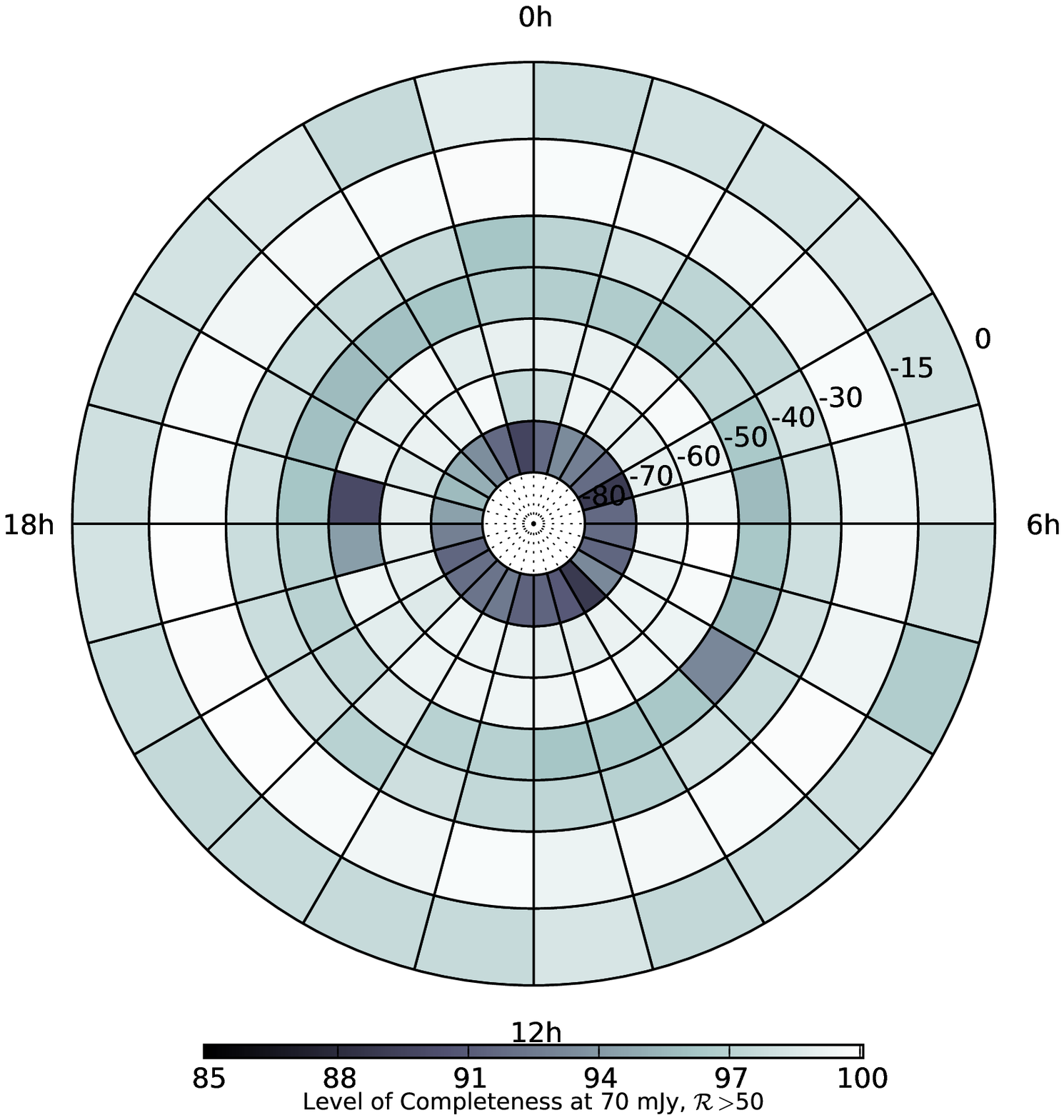,bb=70 135 540 625, width=0.45\linewidth,clip=}
 \epsfig{file=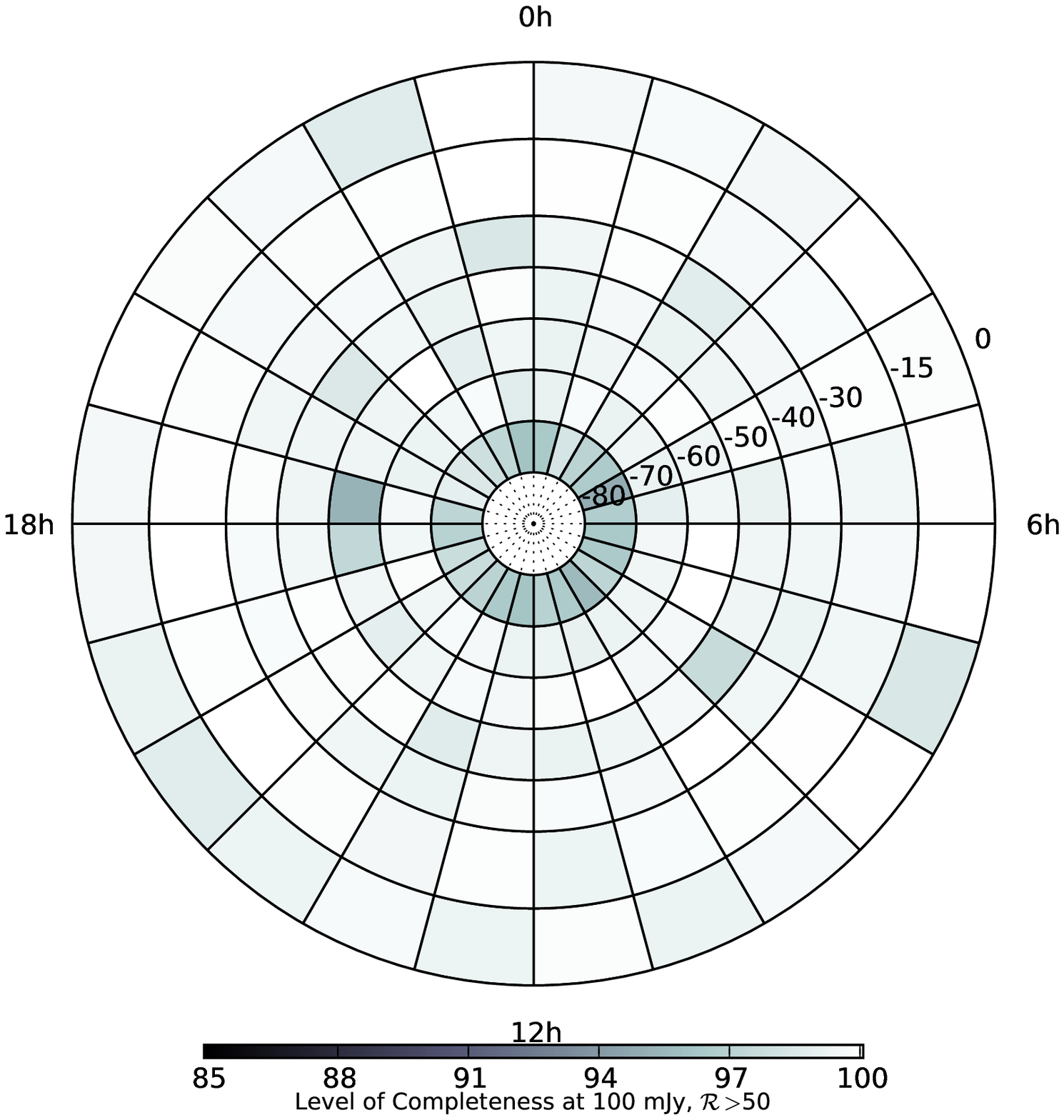,bb=70 135 540 625, width=0.45\linewidth,clip=}
 \caption[The completeness of the survey as measured via the injection and recovery of test sources]{The completeness of the survey as measured via the injection and recovery of test sources, at 70\,mJy ({\bf Left}) and 100\,mJy ({\bf Right}).}
 \label{fig:completeness_radial}
\end{figure}

The major contribution to the degradation of completeness seen in some maps in figure \ref{fig:completeness_radial}, is changes in sky coverage. The maps that have a completeness below the average of their region have a different colour. In each of these cases the lower completeness is due to data that were flagged as poor but was unable to be recovered during patching observations or that was simply not observed due to weather and scheduling constraints. The scanning survey catalog is 98\% complete at 70\,mJy and 92\% complete at 50\,mJy.

\section{Scanning Survey catalog}
The final version of the scanning survey catalog (SSC) is made of from sources with an SNR $>5$ and a reliability of 50\% or greater. Table \ref{tab:surveycat} shows the first page of the scanning survey catalog, which contains a total of 7754 sources. The catalog format is as follows:
\begin{itemize}{}{}
\item Columns  1\&2: The source RA/DEC in J2000 co-ordinates
\item Columns  3\&4: Galactic $(l,b)$
\item Columns  5\&6: Corrected survey flux at 20\,GHz and the map rms noise (mJy)
\item Column 7: Reliability $\mathcal{R}$
\item Column 8: Epoch of scanning observation. Refer to table \ref{tab:obsdates} for the precise dates
\item Column 9: AT20G name, when the source is in the follow-up survey catalog of \citet{murphy_australia_2010}
\end{itemize}
The typical positional errors in $\alpha$ and $\delta$ are $\pm 15 $arc-sec. The flux scale has been corrected to be the same as the follow-up survey catalog.  There may be variations in the individual fluxes due to source variability and the lower signal to noise of the survey maps. The AT20G names are determined by cross matching the follow-up survey catalog (FSC) and the scanning survey catalog (SSC). The nearest scanning survey catalog source within 120 arcsec is considered to be the correct identification.

\section{Summary}
The AT20G survey was made possible via the use of a wide band analog correlator in conjunction with an interleaved fast scanning strategy (or on-the-fly mapping). New programs were created that allowed for the offline calibration of the scanning survey data. The subsequent calibration scheme was accurate enough that the scanning data could be brought together to form convolved maps of the sky from which candidate source lists could be created. The offline data processing techniques allowed artificial sources to be introduced into the data reduction pipeline which allowed for an accurate measurement of the completeness of the scanning survey catalog, and by extension, the completeness of the follow-up survey catalog as described in \citep{murphy_australia_2010}.

The scanning survey catalog presented here has a lower flux limit and is more complete than the follow-up survey catalog, but achieves these goals at the cost of reliability. The follow-up survey catalog is $\sim$100\% reliable, whilst the scanning survey catalog includes some sources with only 50\% reliability. It is expected that the primary use of this catalog will be to provide foreground masks for CMB missions such as Planck \citep{tauber_planck_2005} and the South Pole Telescope \citep{stark_south_1998}, where completeness is more important than accurate fluxes or a high reliability.

\begin{table}[hbt]
 \centering
 \begin{tabular}{ccrrrrccc}
\hline
$\alpha$ & $\delta$ & $l$ & $b$ & $S_{20}$ & map rms & reliability & epoch & AT20G\\
(J2000)  & (J2000)  & \deg  & \deg  & mJy & mJy  & $\mathcal{R}$ & & name \\
 (1)     & (2) & (3) & (4) & (5) & (6) & (7) & (8) &(9)\\
\hline
00:00:02 & -00:22:30 & 96.0 & -60.5 & 58 & 10 & 0.93 & 2007 & \\
00:00:12 & -03:43:45 & 93.1 & -63.6 & 51 & 10 & 0.89 & 2007 & \\
00:00:21 & -32:21:00 & 4.6 & -77.8 & 160 & 12 & 0.88 & 2004 & J000020-322101\\
00:01:06 & -15:51:30 & 74.4 & -73.8 & 306 & 9 & 0.98 & 2006 & J000105-155107\\
00:01:06 & -17:42:00 & 69.4 & -75.0 & 67 & 9 & 0.98 & 2006 & J000106-174126\\
00:01:12 & -00:19:45 & 96.6 & -60.6 & 53 & 10 & 0.84 & 2007 & \\
00:01:18 & -07:47:00 & 89.1 & -67.3 & 182 & 10 & 0.98 & 2007 & J000118-074626\\
00:01:25 & -04:38:30 & 92.8 & -64.5 & 75 & 10 & 0.98 & 2007 & J000124-043759\\
00:01:26 & -06:56:30 & 90.2 & -66.6 & 69 & 10 & 0.98 & 2007 & J000125-065624\\
00:01:41 & -15:41:00 & 75.1 & -73.7 & 48 & 9 & 0.94 & 2006 & \\
00:02:12 & -21:53:30 & 55.3 & -77.6 & 120 & 9 & 0.98 & 2006 & J000212-215309\\
00:02:22 & -14:07:15 & 79.1 & -72.7 & 78 & 10 & 0.98 & 2007 & J000221-140643\\
00:02:31 & -03:31:45 & 94.5 & -63.7 & 81 & 10 & 0.94 & 2007 & J000230-033140\\
00:02:50 & -21:14:30 & 58.3 & -77.5 & 142 & 9 & 0.98 & 2006 & J000249-211419\\
00:02:53 & -59:48:15 & 313.9 & -56.3 & 104 & 9 & 0.96 & 2005 & J000252-594814\\
00:02:54 & -56:21:00 & 316.2 & -59.5 & 113 & 9 & 0.98 & 2005 & J000253-562110\\
00:03:03 & -55:29:45 & 316.8 & -60.3 & 71 & 9 & 0.77 & 2005 & J000303-553007\\
00:03:10 & -54:44:45 & 317.4 & -61.0 & 151 & 9 & 0.98 & 2005 & J000311-544516\\
00:03:13 & -59:05:45 & 314.3 & -57.0 & 66 & 9 & 0.90 & 2005 & J000313-590547\\
00:03:16 & -19:42:15 & 64.3 & -76.7 & 100 & 9 & 0.98 & 2006 & J000316-194150\\
00:03:19 & -19:27:30 & 65.2 & -76.6 & 129 & 9 & 0.98 & 2006 & \\
00:03:22 & -17:27:30 & 71.5 & -75.3 & 420 & 9 & 0.98 & 2006 & J000322-172711\\
00:03:27 & -15:47:15 & 76.0 & -74.1 & 143 & 9 & 0.98 & 2006 & J000327-154705\\
00:03:45 & -11:09:00 & 85.6 & -70.5 & 53 & 10 & 0.93 & 2007 & \\
00:03:48 & -23:30:00 & 48.9 & -78.6 & 70 & 9 & 0.93 & 2006 & \\
00:03:57 & -53:43:00 & 318.0 & -62.0 & 48 & 9 & 0.70 & 2005 & \\
00:04:05 & -11:49:15 & 84.7 & -71.1 & 618 & 10 & 0.98 & 2007 & J000404-114858\\
00:04:07 & -43:45:15 & 329.7 & -70.8 & 223 & 12 & 0.98 & 2004 & J000407-434510\\
00:04:14 & -52:54:45 & 318.6 & -62.8 & 73 & 9 & 0.91 & 2005 & J000413-525458\\
00:04:15 & -07:08:45 & 91.5 & -67.1 & 51 & 10 & 0.82 & 2007 & \\
00:04:35 & -47:36:15 & 324.0 & -67.6 & 995 & 12 & 0.98 & 2004 & J000435-473619\\
00:04:47 & -66:16:00 & 310.2 & -50.2 & 68 & 12 & 0.93 & 2005 & \\
00:05:07 & -01:33:15 & 97.5 & -62.1 & 102 & 10 & 0.98 & 2007 & J000507-013244\\
00:05:07 & -34:45:45 & 352.5 & -77.5 & 125 & 12 & 0.94 & 2004 & J000505-344549\\
00:05:18 & -16:48:30 & 74.6 & -75.2 & 156 & 9 & 0.98 & 2006 & J000518-164804\\
00:05:24 & -57:52:00 & 314.6 & -58.2 & 47 & 9 & 0.74 & 2005 & \\
00:05:58 & -56:28:30 & 315.3 & -59.5 & 182 & 9 & 0.98 & 2005 & J000558-562828\\
00:05:59 & -06:12:00 & 93.6 & -66.4 & 63 & 10 & 0.70 & 2007 & \\
00:06:02 & -29:56:15 & 14.5 & -79.6 & 80 & 9 & 0.98 & 2006 & J000601-295549\\
00:06:02 & -42:34:15 & 330.9 & -72.0 & 118 & 12 & 0.98 & 2004 & J000601-423439\\
00:06:02 & -31:32:45 & 6.2 & -79.2 & 86 & 12 & 0.89 & 2004 & J000600-313215\\
00:06:14 & -06:24:00 & 93.5 & -66.7 & 2287 & 10 & 0.98 & 2007 & J000613-062334\\
00:06:19 & -42:45:30 & 330.4 & -71.9 & 147 & 12 & 0.90 & 2004 & J000619-424518\\
00:06:23 & -00:05:00 & 99.3 & -60.9 & 432 & 10 & 0.98 & 2007 & J000622-000423\\
00:06:30 & -73:12:00 & 307.4 & -43.5 & 65 & 10 & 0.62 & 2005 & J000635-731144\\
00:07:13 & -40:23:30 & 334.7 & -73.9 & 86 & 12 & 0.77 & 2004 & J000713-402337\\
00:07:20 & -61:13:00 & 312.2 & -55.1 & 193 & 12 & 0.98 & 2005 & J000720-611306\\
00:08:00 & -23:40:00 & 50.0 & -79.6 & 154 & 9 & 0.98 & 2006 & J000800-233918\\
00:08:02 & -52:43:30 & 317.6 & -63.2 & 107 & 9 & 0.98 & 2005 & J000801-524339\\
00:08:10 & -39:45:15 & 335.6 & -74.5 & 163 & 12 & 0.98 & 2004 & J000809-394522\\
00:08:26 & -25:59:15 & 37.3 & -80.3 & 109 & 9 & 0.98 & 2006 & J000826-255911\\

\hline
\end{tabular}
\caption{The first page of the scanning survey catalog. A pre-publication version of the entire catalog is available at www.physics.usyd.edu.au/$\sim$hancock/files/SurveyCatV1.1}
\label{tab:surveycat}
\end{table}

\clearpage
\bibliographystyle{spbasic}
\bibliography{Hancock2011}

\end{document}